\documentclass[fleqn,usenatbib]{mnras}

\usepackage{newtxtext,newtxmath}
\usepackage[T1]{fontenc}

\DeclareRobustCommand{\VAN}[3]{#2}
\let\VANthebibliography\thebibliography
\def\thebibliography{\DeclareRobustCommand{\VAN}[3]{##3}\VANthebibliography}

\usepackage{graphicx}	% Including figure files
\usepackage{amsmath}	% Advanced maths commands

\title[]{Detection of episodic Ultra-fast Outflows in the Low-luminosity quasar Mrk\,205 using \emph{XMM}-Newton data}

\author[Victoria-Ceballos et al.]{
C\'esar Ivan Victoria-Ceballos,$^{1}$\thanks{E-mail: cvictoria@astro.unam.mx (KTS)}
Anna Lia Longinotti,$^{1}$
Yair Krongold,$^{1}$
Giorgio Lanzuisi,$^{2}$
\newauthor Miriam Gudiño,$^{3}$ Armando Lara-DI,$^{4}$
\\
$^{1}$Instituto de Astronomía, (IA-UNAM), Circuito Exterior, Ciudad Universitaria, Ciudad de México 04510, México\\
$^{2}$INAF – Osservatorio di Astrofisica e Scienza dello Spazio di Bologna, Via Gobetti, 93/3, 40129 Bologna, Italy\\
$^{3}$Instituto Nacional de Astrofísica, Óptica y Electrónica, Luis Enrique Erro $\#$1, Tonantzintla, Puebla 72840, México \\
$^{4}$Instituto de Ciencias Nucleares, Universidad Nacional Autónoma de México, 04510 Ciudad de México, México \\
}

\date{Accepted XXX. Received YYY; in original form ZZZ}

\pubyear{\the\year{}}

\begin{document}
\label{firstpage}
\pagerange{\pageref{firstpage}--\pageref{lastpage}}
\maketitle

\begin{abstract}
Ultra-fast outflows (UFOs) from Active Galactic Nuclei (AGN) are winds which reach velocities up to 0.3c, and show mass outflow rates of 0.01-1 $M_{\odot} yr^{-1}$ and kinetic energies of $10^{42-45}$ erg/s. Their extreme properties allow UFOs to carry substantial kinetic power, sufficient to drive AGN feedback by injecting energy and momentum into the interstellar medium, regulating the host galaxy evolution and the supermassive black hole growth. 
In this work, we present a detailed analysis of \emph{XMM}-Newton X-ray spectra of the low-luminosity quasar Mrk205. We find evidence of multi-component ultra-fast outflows, traced by two components detected in the high-resolution data, showing velocities in the range of $\sim$0.07c to $\sim$0.1c. In addition, we marginally detected an ultra-fast outflow feature in the Fe\,K band. According to our analysis, Mrk\,205 shows episodes of wind launches, rather than persistent winds. We further discuss whether the energy carried by the UFOs in Mrk\,205 is capable of producing significant feedback on its host galaxy.
\end{abstract}


%%%%%%%%%%%%%%%%%%%%%%%%%%%%%%%%%%%%%%%%%%%%%%%%%%

%%%%%%%%%%%%%%%%% BODY OF PAPER %%%%%%%%%%%%%%%%%%
\section{Introduction}

Active Galactic Nuclei (AGN) are powered by accretion of matter onto the supermassive black hole residing at the center of their host galaxy. Efficient accretion processes can launch powerful accretion disk winds, producing highly ionized gas observable in the X-ray band. Among these disk winds, Ultra-fast Outflows (UFOs) are the most extreme manifestation, characterized by high ionization ($log[\xi(erg cm s^{-1}$)]=3-6), high column densities ($log[N_{H} (cm^{-2}=22-24)]$), and reach high velocities of $v_{out} > 10^{4} km/s$ \citep[][]{Laha21}. Their inferred mass outflow rates (0.01-1 $M_{\odot}$/yr), and kinetic energies ($10^{42-45}$ erg/s) indicate that they can contribute to regulate host galaxy properties by  starting a process that ultimately leads to alter star formation and galaxy evolution \citep{DiMatteo05, Hopkins10, King15}. 

Ultra-fast outflows are primarily identified through blue-shifted absorption features in the X-ray spectra. They are detected mainly as Fe\,XXV He$\alpha$ ($\sim$6.67 keV) and Fe\,XXVI Ly$\alpha$ ($\sim$6.97 keV) resonant absorption lines in the Fe\,K band, between 7–10 keV \citep{Tombesi10, Gofford13}. However, UFO features above 10 keV have also been detected, revealing  higher speed components of the wind \citep{Matzeu23, Luminari23}. Recent observations have revealed that UFOs often exhibit multiple highly ionized components with different velocities and ionization states, indicating a complex and stratified wind structure \citep{Parker20, Xiang25, Xu25}. In addition, absorption features from lighter ions such as O, Ne, Mg, and Si are detected in the soft X-ray band, providing critical insight into the multiphase structure of these winds, linking the most ionized relativistic components to lower-ionization gas phases \citep{Reeves09, Matzeu17, Chartas21}.

While the presence of ultra-fast outflows is frequently associated with moderate-to-high accreting sources \citep{Longinotti15, Parker17}, where radiation pressure is key to accelerating winds to sub-relativistic velocities, ultra-fast outflows have also been detected in low accretion sources \citep[eg.][]{Matzeu23}, suggesting that their occurrence is not restricted to a narrow interval of Eddington ratios. Although radiative driving is expected to play a major role in high accreting sources, additional mechanisms such as magneto-hydrodynamic (MHD) processes may contribute to wind launching across different accretion regimes \citep{Fukumura17,Xu25}. 

Mrk\,205 is a nearby low-luminosity quasar at redshift z=0.070846, with black hole mass of $M_{BH} \sim 2.1\times10^8 M_{\odot}$, bolometric luminosity of $L_{BOL}$ $\sim 7.4\times10^{44}$ erg $s^{-1}$, and Eddington ratio of $\lambda_{Edd}$ $\sim$ 0.03 \citep{Salome23}, which has previously been studied in the context of UFOs by \cite{Tombesi10}. They reported a relativistic outflow with velocity of $\sim$ 0.1 c detected in the  FeK$\alpha$ band in one \emph{XMM}-Newton CCD spectrum. 
More recently, \cite{Laha19} reported on the presence of partially ionized absorption in \emph{XMM}-Newton and \emph{Suzaku} data of Mrk\,205, with no confirmation of the previously reported UFO feature, without clear evidence for disk reflection.

In this work, we present a detailed analysis of the high- and medium-resolution data of Mrk\,205 using available \emph{XMM}-Newton data, in order to investigate the presence and properties of the UFOs in this source. We describe the data reduction in Section\,\ref{sec: Data reduction}. The X-ray spectral fitting procedure and the results of the spectral fitting are described in Section\,\ref{sec: Spectral fitting}. We discuss the results in Section\,\ref{sec:Discussion}.

\begin{figure}
\begin{center}
\includegraphics[width=1.0\columnwidth]{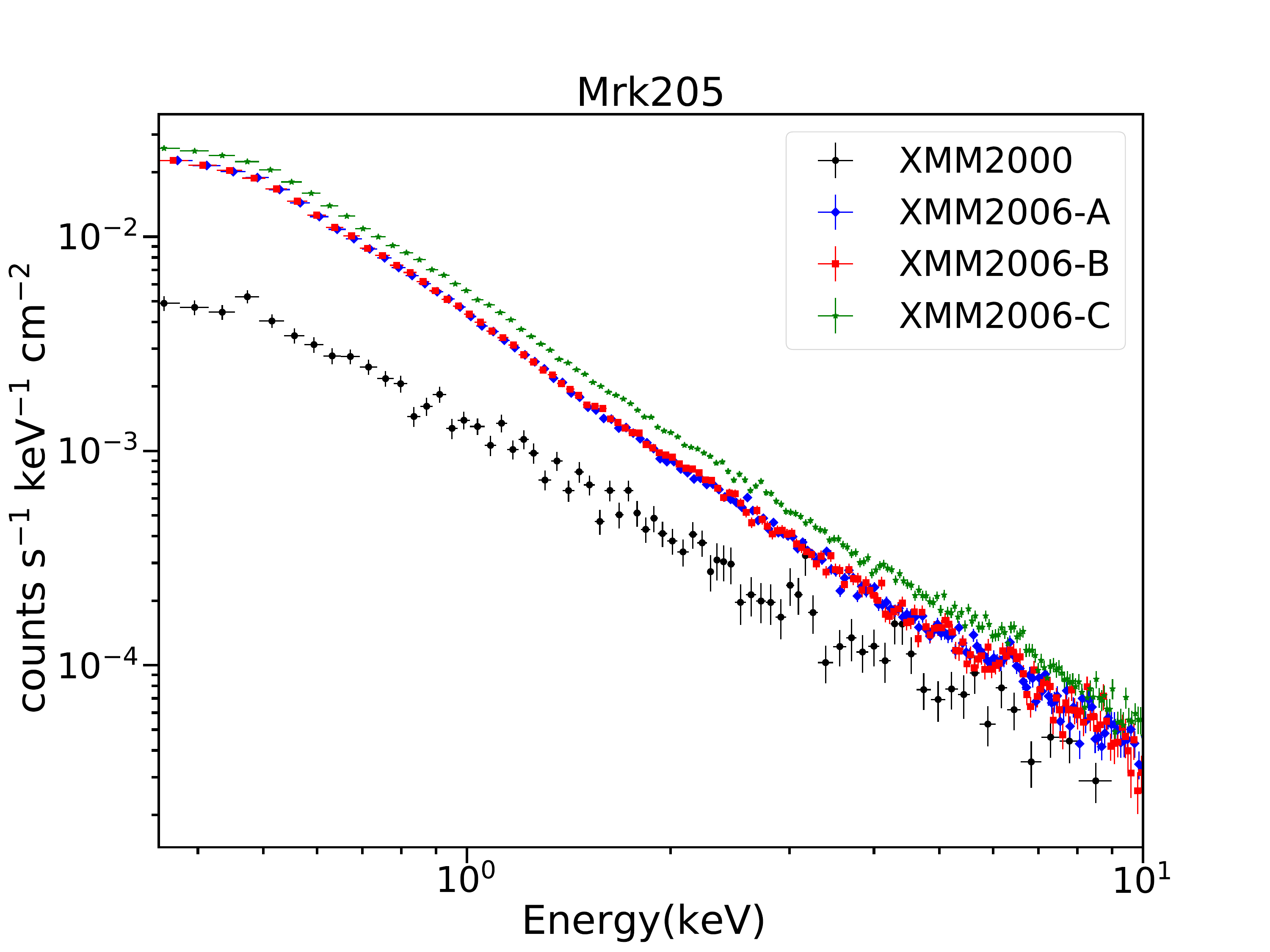}\\
\caption{\emph{XMM}-Newton spectra of Mrk\,205. Black points correspond to the \emph{XMM}-Newton observation from 2000, labeled as XMM2000. Blue, red, and green points correspond to observations 0401240201 (XMM2006-A), 0401240301 (XMM2006-B), and 0401240501 (XMM2006-C) from \emph{XMM}-Newton 2006 observations, respectively.}
\label{fig:Mrk205_observations}
\end{center}
\end{figure}

%%%%%%%%%%%%%%%%%%%%%%%%%%%%%%%%%%%%%%%%%%%%%%%%%%%%%%%%%%%%%%%%%%%%%%%%%%%
%%%%%%%%%%%%%%%%%%%%%%%%%%%%%%%%%%%%%%%%%%%%%%%%%%%%%%%%%%%%%%%%%%%%%%%%%%%
%%%%%%%%%%%%%%%%%%%%%%%%%%%%%%%%%%%%%%%%%%%%%%%%%%%%%%%%%%%%%%%%%%%%%%%%%%%
\begin{table*}
%\tiny
%\scriptsize 
\renewcommand{\tabcolsep}{0.18cm}
\begin{center}
\begin{tabular}{ccccccccccc} \hline
Obs. ID & Obs. date & Exp. & \multicolumn{2}{c}{0.2-10 keV band} & \multicolumn{2}{c}{0.2-2 keV band} & \multicolumn{2}{c}{2-10 keV band} \\
& & (ks) & $\Gamma$ & Flux & $\Gamma$ & Flux & $\Gamma$ & Flux \\ \hline
0124110101 & 2000-05-07 & 66.2 & $\rm{1.97\pm0.01}$ & $\rm{1.145\pm0.005}\times10^{-11}$ & $\rm{2.06\pm0.01}$ & $\rm{6.65\pm0.03}\times10^{-12}$ & $\rm{1.69\pm0.03}$ & $\rm{5.59\pm0.06}\times10^{-12}$ \\
0401240201 & 2006-10-18 & 32.9 & $\rm{2.42\pm0.01}$ & $\rm{2.915\pm0.007}\times10^{-11}$ & $\rm{2.53\pm0.01}$ & $\rm{2.29\pm0.01}\times10^{-11}$ & $\rm{1.98\pm0.02}$ & $\rm{8.89\pm0.06}\times10^{-12}$ \\
0401240301 & 2006-10-20 & 58.9 & $\rm{2.41\pm0.01}$ & $\rm{2.929\pm0.007}\times10^{-11}$ & $\rm{2.51\pm0.01}$ & $\rm{2.29\pm0.01}\times10^{-11}$ & $\rm{2.01\pm0.02}$ & $\rm{8.86\pm0.06}\times10^{-12}$ \\
0401240501 & 2006-10-22 & 59.3 & $\rm{2.31\pm0.01}$ & $\rm{3.606\pm0.008}\times10^{-11}$ & $\rm{2.40\pm0.01}$ & $\rm{2.65\pm0.01}\times10^{-11}$ & $\rm{1.98\pm0.01}$ & $\rm{1.20\pm0.06}\times10^{-11}$ \\
\hline
\end{tabular}
\end{center}
\caption{Observational parameters of the 4 \emph{XMM}-Newton data of Mrk\,205. Cols.\,1, 2, and 3 show the observation ID, the date, and the exposure time of the observation, respectively. Cols.\,4-5, 6-7, and 8-9 show the photon index and the unabsorbed flux (in ${\rm erg\,cm^{-2}\,s^{-1}}$) of the source in the 0.2-10 keV, 0.2-2 keV, and 2-10\,keV energy bands, respectively.}
\label{tab:Observational_parameters}
\end{table*}
%%%%%%%%%%%%%%%%%%%%%%%%%%%%%%%%%%%%%%%%%%%%%%%%%%%%%%%%%%%%%%%%%%%%%%%%%%%
%%%%%%%%%%%%%%%%%%%%%%%%%%%%%%%%%%%%%%%%%%%%%%%%%%%%%%%%%%%%%%%%%%%%%%%%%%%
%%%%%%%%%%%%%%%%%%%%%%%%%%%%%%%%%%%%%%%%%%%%%%%%%%%%%%%%%%%%%%%%%%%%%%%%%%%

\section{Data reduction} \label{sec: Data reduction}

Mrk\,205 was observed by \emph{XMM}-Newton 4 times (see Tab.\,\ref{tab:Observational_parameters}). Spectra from the Reflection Grating Spectrometer \citep[RGS;][]{denHerder01} and European Photon Imaging Camera \citep[EPIC pn;][]{Struder01} from \emph{XMM}-Newton were processed by the standard  System Analysis Software (SAS)\footnote{https://www.cosmos.esa.int/web/xmm-newton/sas} tools \emph{rgsproc} and \emph{epproc}, respectively. In order to maximize the signal-to-noise ratio, RGS and EPIC pn spectra of 2006 observations were combined in a single spectrum by the SAS tools \emph{rgscombine} and \emph{mathpha}, respectively. We show in Fig.\,\ref{fig:Mrk205_observations}  the EPIC-pn spectra by \emph{XMM}-Newton. Note that, the observations analyzed by \cite{Tombesi10} was the XMM2000, XMM2006-A, and XMM2006-C (see Sec.\,\ref{sec: Discussion CCD spectra}).  
To carry out the RGS and the CCD spectral analysis we applied the C-statistic and the $\chi^2$ statistic, respectively. We used circular regions with 40 arcsec radii to extract the spectra (corresponding to an encircled energy fraction of $\sim$90\%). The background events were selected from a source-free circular region with 40 arcsec radii on the same CCD as the source. For the RGS, we binned the spectra o a minimum of one count per bin, in order to preserve the intrinsic spectral resolution. For the CCD, we grouped the data using the binning scheme of \cite{Kaastra16}, which adapts the bin size according to the instrumental resolution and the number of counts. To determine the confidence intervals of the parameters, we estimate the ${\rm 1\sigma}$ errors.

In order to maximize the signal-to-noise ratio and combine the contemporaneous observations, we determined the flux variability and the lack of spectral changes of the source by computing the photon index and the flux in the 0.2-2, the 2-10, and the 0.2-10\,keV energy bands. We found a minor difference in the photon index and flux in the three bands (see Cols.\,4-9 in Tab.\,\ref{tab:Observational_parameters}). Given the negligible variability of the continuum flux and spectral shape, and given the proximity in time, we decided to  carry out the spectral analysis on the stacked 3 spectra from the consecutive 2006 observations (refereed to as XMM2006) and on the spectra from 2000 (referred to as XMM2000).
A consistency check on the individual observations of XMM2006 is provided in Appendix \ref{appendix:XMM-Newton observations from 2006}.

\section{Spectral fitting} \label{sec: Spectral fitting}

\subsection{Spectral fitting of the RGS data} \label{sec: Spectral fitting of the RGS data}  

We performed our spectral fitting using the {\sc XSPEC} \footnote{http://heasarc.gsfc.nasa.gov/docs/xanadu/xspec/} fitting package \citep{Arnaud96}, which is a command-driven, interactive, spectral-fitting program within the HEASOFT \footnote{https://heasarc.gsfc.nasa.gov} software.

We started by fitting the X-ray continuum of the \emph{XMM}-Newton observations with a power law modified by the Galactic absorption \citep[fixed to the HI gas measured by][]{Kalberla05}.          

In order to search for the absorption features in the spectra by the UFOs, we added a narrow ($\sigma$=0.1 eV) Gaussian line with negative intensity and free position and searched for absorption features using the \emph{steppar} command in XSPEC, in the 7-37\,\r{A} range. The method consists of computing the $\Delta\chi^2$ deviations from the best-fitting model, and calculating the contour plots of the energy-intensity plane.

\begin{figure}
\begin{center}
\includegraphics[width=1.0\columnwidth]{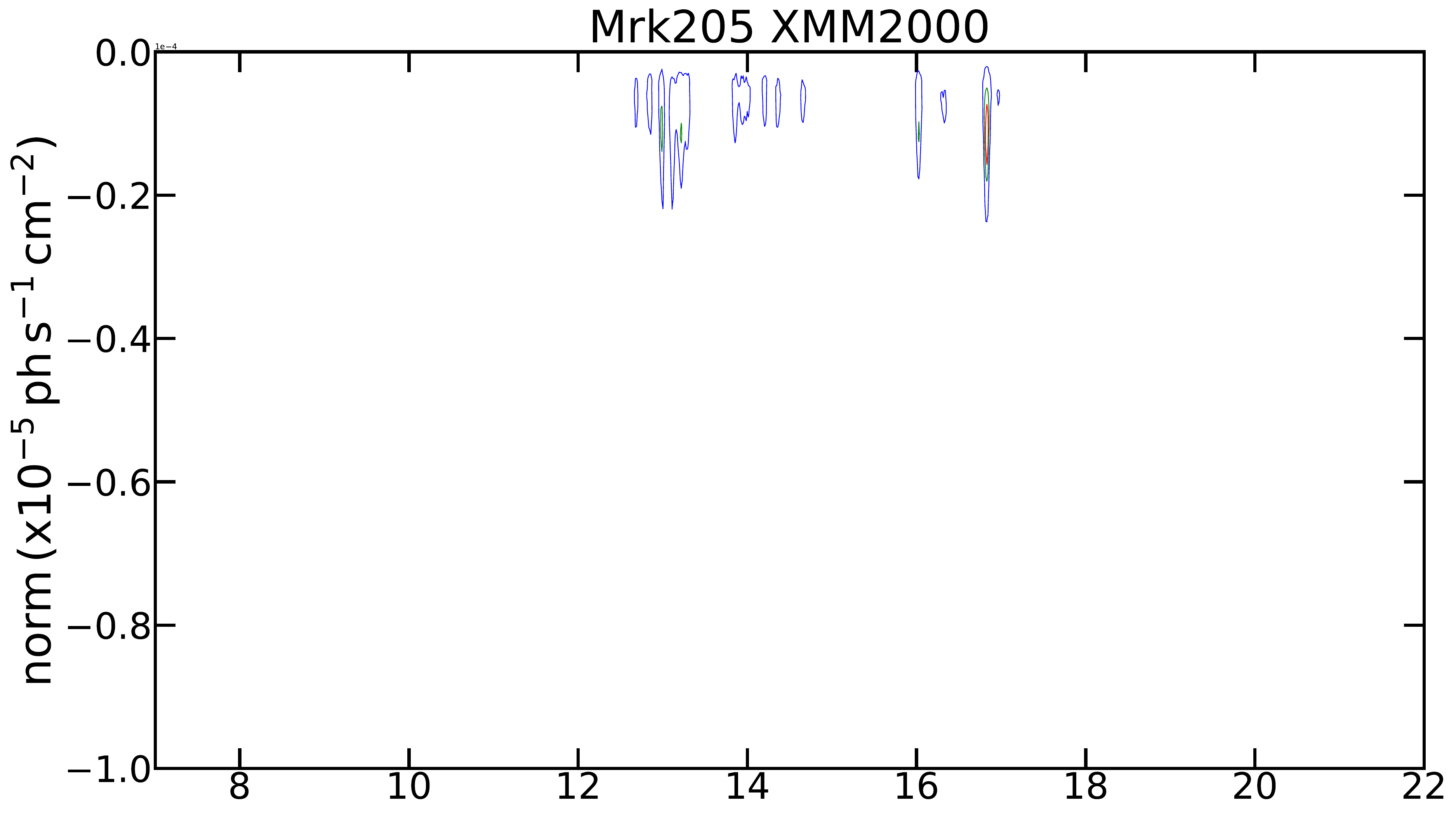}\\
\includegraphics[width=1.0\columnwidth]{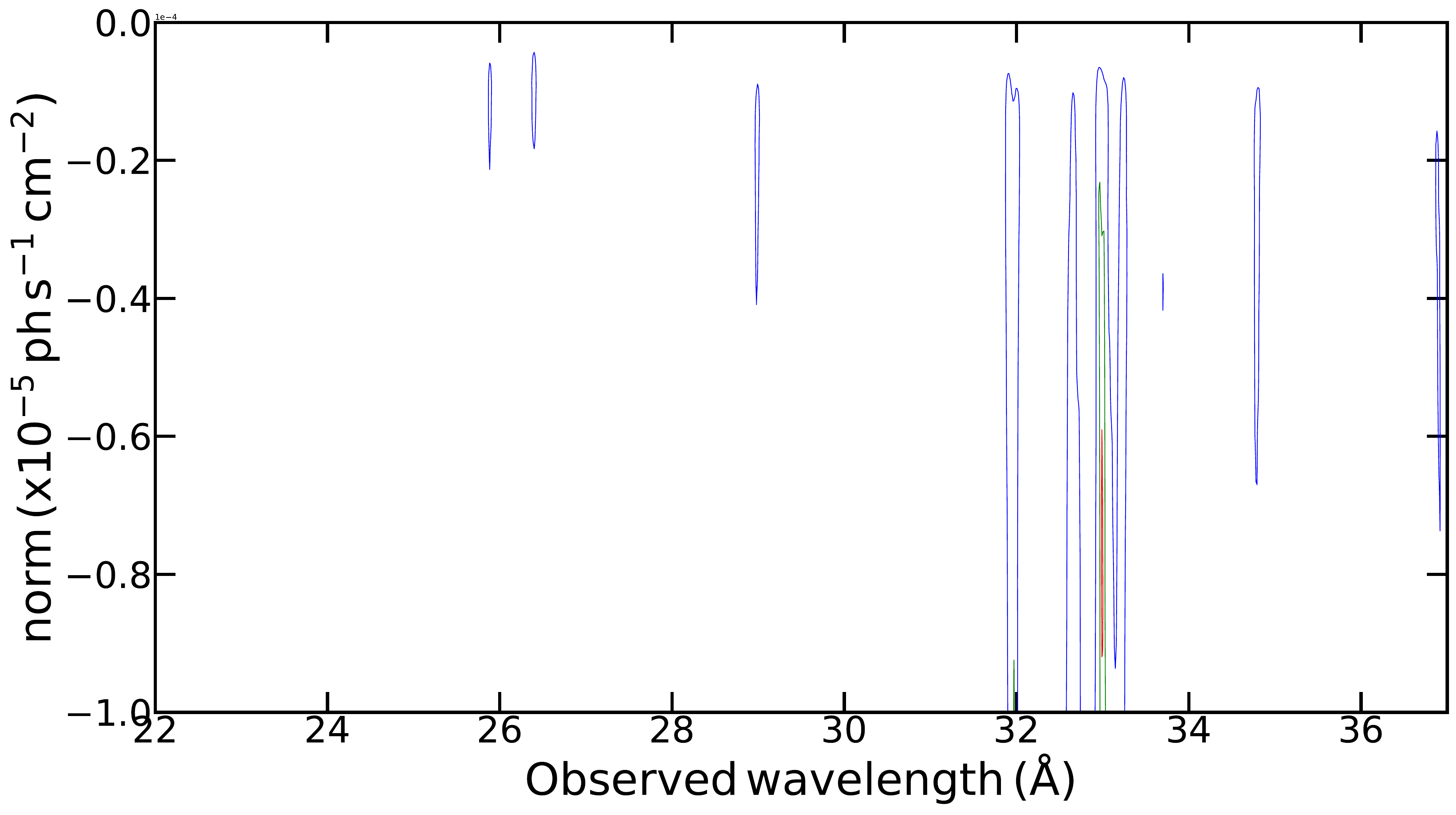}
\caption{Two-dimensional $\Delta\chi^2$ contours for the resulting line search performed in the RGS spectra of XMM2000. Red, green, and blue contours correspond to confidence levels of 68$\%$, 90$\%$, and 99$\%$, respectively.}
\label{fig:Abs gauss contours XMM2000}
\end{center}
\end{figure}

\begin{figure}
\begin{center}
\includegraphics[width=1.0\columnwidth]{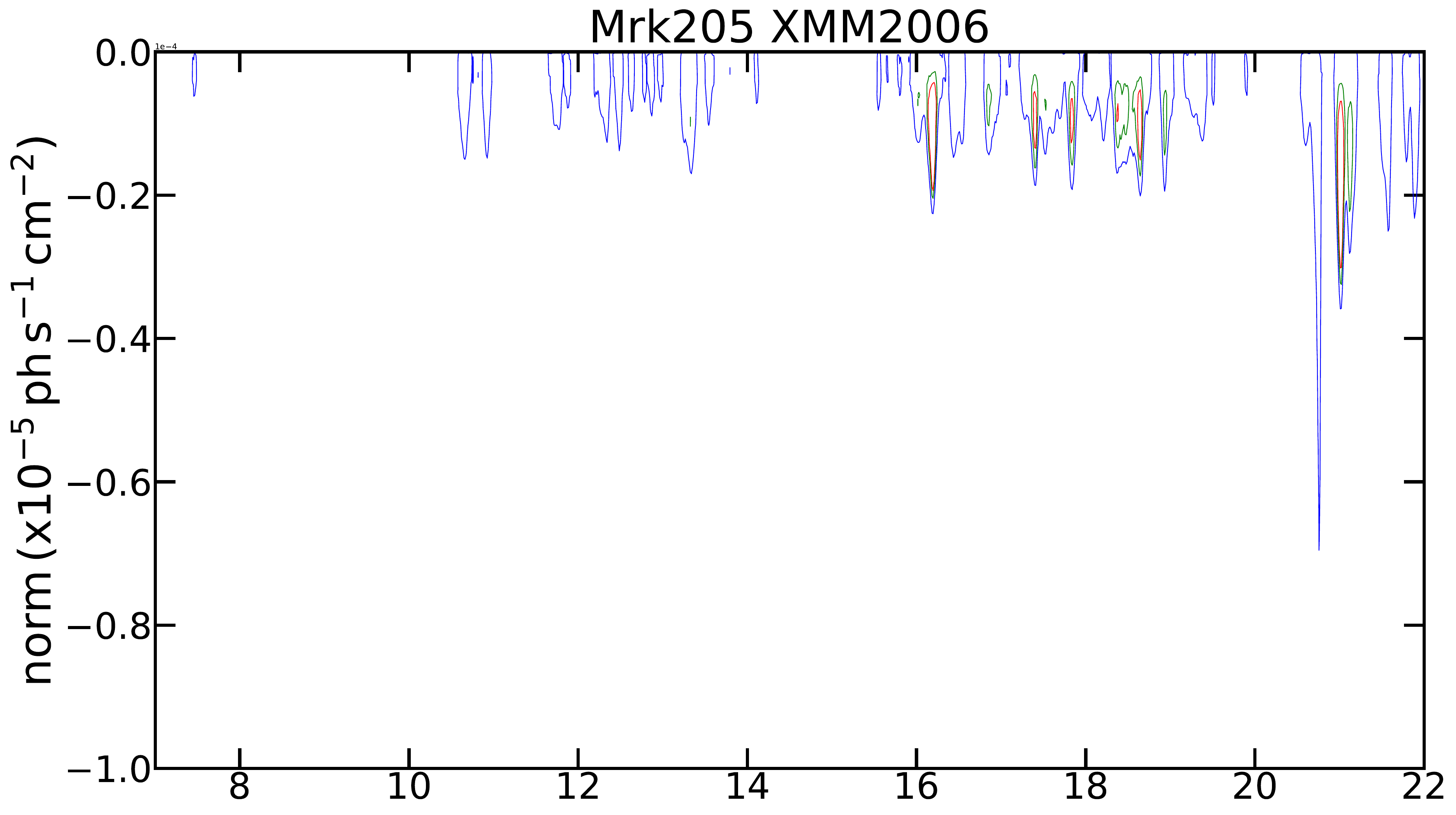}\\
\includegraphics[width=1.0\columnwidth]{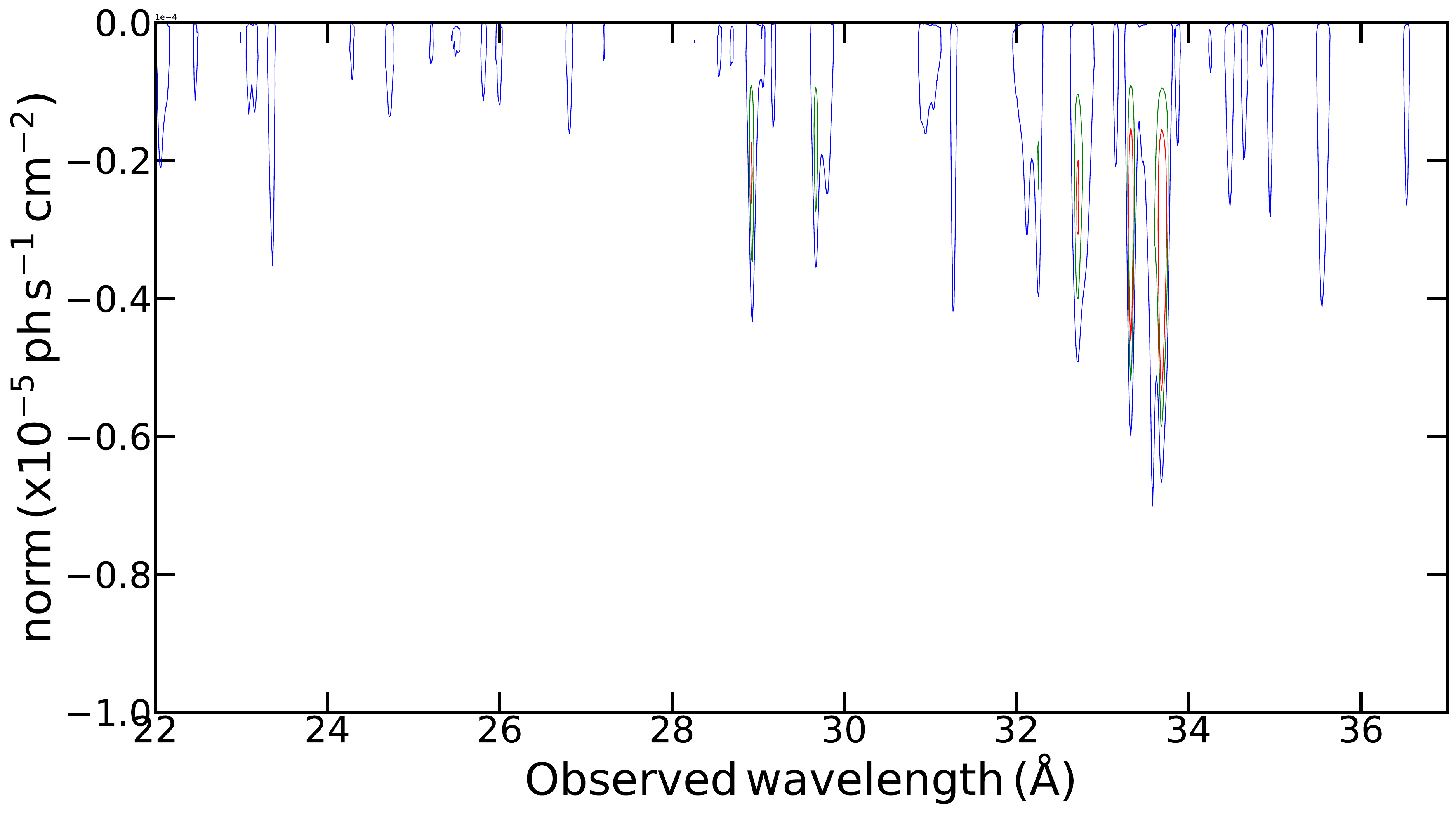}
\caption{Two-dimensional $\Delta\chi^2$ contours for the resulting line search performed in the RGS spectra of XMM2006. Red, green, and blue contours correspond to confidence levels of 68$\%$, 90$\%$, and 99$\%$, respectively. }
\label{fig:Abs gauss contours XMM2006}
\end{center}
\end{figure}

We show in Fig.\,\ref{fig:Abs gauss contours XMM2000} and Fig.\,\ref{fig:Abs gauss contours XMM2006} the contour plots of the absorption lines search. The levels are $\Delta\chi^2$=2.3, 4.61, and 9.21, which correspond to confidence levels of 68$\%$, 90$\%$, and 99$\%$, respectively. Tab.\,\ref{tab:Abs_gauss} shows  the list of the absorption lines detected. The results show that additional absorption is required by the data; thus, we proceeded to model it through a self-consistent model.

%%%%%%%%%%%%%%%%%%%%%%%%%%%%%%%%%%%%%%%%%%%%%%%%%%%%%%%%%%%%%%%%%%%%%%%%%%%
%%%%%%%%%%%%%%%%%%%%%%%%%%%%%%%%%%%%%%%%%%%%%%%%%%%%%%%%%%%%%%%%%%%%%%%%%%%
%%%%%%%%%%%%%%%%%%%%%%%%%%%%%%%%%%%%%%%%%%%%%%%%%%%%%%%%%%%%%%%%%%%%%%%%%%%
\begin{table}
%\tiny
%\scriptsize 
\renewcommand{\tabcolsep}{0.05cm}
\begin{center}
\begin{tabular}{ccccccccc} \hline
Obs. & Obs. $\lambda$ & Intensity & $\Delta$C-stat & Sign. & Line ID & Component \\
& (\r{A}) & ($\rm{10^{-5}\,ph\,cm^{-2}\,s^{-1}}$) & & ($\sigma$) & \\ 
\hline
XMM2000 & $\rm{16.83\pm^{0.04}_{0.01}}$ & $\rm{-1.17\pm^{0.32}_{0.26}}$ & 12 & 3.66 & ? & ? \\
& $\rm{16.03\pm^{0.01}_{0.01}}$ & $\rm{-1.03\pm^{0.33}_{0.32}}$ & 8 & 3.12 & ? & ? \\
& $\rm{32.99\pm^{0.02}_{0.01}}$ & $\rm{-7.54\pm^{2.43}_{2.95}}$ & 10 & 3.10 & ? & ? \\
& $\rm{13.22\pm^{0.04}_{0.04}}$ & $\rm{-1.06\pm^{0.35}_{0.33}}$ & 8 & 3.03 & Fe\,XIX & UFO\,2 \\
& $\rm{12.98\pm^{0.01}_{0.01}}$ & $\rm{-1.13\pm^{0.38}_{0.32}}$ & 9 & 2.97 & ? & ? \\
& $\rm{29.33\pm^{0.02}_{0.10}}$ & $\rm{-1.53\pm^{0.68}_{0.64}}$ & 6 & 2.25 & Ne\,III & UF\,1 \\
& $\rm{31.91\pm^{0.04}_{0.02}}$ & $\rm{-2.60\pm^{1.30}_{3.26}}$ & 6 & 2 & ? & ? \\
\hline
XMM2006 & $\rm{16.20\pm^{0.02}_{0.02}}$ & $\rm{-1.20\pm^{0.32}_{0.32}}$ & 13 & 3.75 & ? & ? \\
& $\rm{21.02\pm^{0.06}_{0.01}}$ & $\rm{-1.94\pm^{0.52}_{0.50}}$ & 12 & 3.73 & O VII & UFO\,2 \\
& $\rm{33.68\pm^{0.02}_{0.02}}$ & $\rm{-3.52\pm^{1.04}_{1.02}}$ & 10 & 3.38 & C\,VI & UFO\,1 \\
& $\rm{17.40\pm^{0.01}_{0.02}}$ & $\rm{-0.99\pm^{0.31}_{0.31}}$ & 9 & 3.19 & ? & ? \\
& $\rm{18.64\pm^{0.02}_{0.02}}$ & $\rm{-1.05\pm^{0.33}_{0.32}}$ & 8 & 3.18 & O VII? & UFO\,1 \\ 
& $\rm{17.84\pm^{0.02}_{0.02}}$ & $\rm{-1.00\pm^{0.34}_{0.33}}$ & 8 & 2.94 & ? & ? \\
& $\rm{28.92\pm^{0.02}_{0.02}}$ & $\rm{-2.26\pm^{0.81}_{0.77}}$ & 7 & 2.79 & ? & ? \\
& $\rm{32.71\pm^{0.02}_{0.08}}$ & $\rm{-2.57\pm^{0.92}_{0.89}}$ & 7 & 2.79 & ? & ? \\
& $\rm{18.38\pm^{0.05}_{0.02}}$ & $\rm{-0.89\pm^{0.32}_{0.31}}$ & 7 & 2.78 & O VIII? & UFO\,2 \\ 
& $\rm{18.94\pm^{0.02}_{0.02}}$ & $\rm{-1.01\pm^{0.39}_{0.39}}$ & 8 & 2.59 & O VIII & UFO\,1 \\ 
\hline
\end{tabular}
\end{center}
\caption{List of absorption lines individually detected in the \emph{XMM}-Newton observations of Mrk\,205 in the 7-37\,\r{A} band.}
\label{tab:Abs_gauss}
\end{table}
%%%%%%%%%%%%%%%%%%%%%%%%%%%%%%%%%%%%%%%%%%%%%%%%%%%%%%%%%%%%%%%%%%%%%%%%%%%
%%%%%%%%%%%%%%%%%%%%%%%%%%%%%%%%%%%%%%%%%%%%%%%%%%%%%%%%%%%%%%%%%%%%%%%%%%%
%%%%%%%%%%%%%%%%%%%%%%%%%%%%%%%%%%%%%%%%%%%%%%%%%%%%%%%%%%%%%%%%%%%%%%%%%%%

%%%%%%%%%%%%%%%%%%%%%%%%%%%%%%%%%%%%%%%%%%%%%%%%%%%%%%%%%%%%%%%%%%%%%%%%%%%
%%%%%%%%%%%%%%%%%%%%%%%%%%%%%%%%%%%%%%%%%%%%%%%%%%%%%%%%%%%%%%%%%%%%%%%%%%%
%%%%%%%%%%%%%%%%%%%%%%%%%%%%%%%%%%%%%%%%%%%%%%%%%%%%%%%%%%%%%%%%%%%%%%%%%%%
%%%%%%%%%%%%%%%%%%%%%%%%%%%%%%%%%%%%%%%%%%%%%%%%%%%%%%%%%%%%%%%%%%%%%%%%%%%
%%%%%%%%%%%%%%%%%%%%%%%%%%%%%%%%%%%%%%%%%%%%%%%%%%%%%%%%%%%%%%%%%%%%%%%%%%%
%%%%%%%%%%%%%%%%%%%%%%%%%%%%%%%%%%%%%%%%%%%%%%%%%%%%%%%%%%%%%%%%%%%%%%%%%%%

%%%%%%%%%%%%%%%%%%%%%%%%%%%%%%%%%%%%%%%%%%%%%%%%%%%%%%%%%%%%%%%%%%%%%%%%%%%
%%%%%%%%%%%%%%%%%%%%%%%%%%%%%%%%%%%%%%%%%%%%%%%%%%%%%%%%%%%%%%%%%%%%%%%%%%%
%%%%%%%%%%%%%%%%%%%%%%%%%%%%%%%%%%%%%%%%%%%%%%%%%%%%%%%%%%%%%%%%%%%%%%%%%%%
%%%%%%%%%%%%%%%%%%%%%%%%%%%%%%%%%%%%%%%%%%%%%%%%%%%%%%%%%%%%%%%%%%%%%%%%%%%
%%%%%%%%%%%%%%%%%%%%%%%%%%%%%%%%%%%%%%%%%%%%%%%%%%%%%%%%%%%%%%%%%%%%%%%%%%%
%%%%%%%%%%%%%%%%%%%%%%%%%%%%%%%%%%%%%%%%%%%%%%%%%%%%%%%%%%%%%%%%%%%%%%%%%%%

%%%%%%%%%%%%%%%%%%%%%%%%%%%%%%%%%%%%%%%%%%%%%%%%%%%%%%%%%%%%%%%%%%%%%%%%%%%
%%%%%%%%%%%%%%%%%%%%%%%%%%%%%%%%%%%%%%%%%%%%%%%%%%%%%%%%%%%%%%%%%%%%%%%%%%%
%%%%%%%%%%%%%%%%%%%%%%%%%%%%%%%%%%%%%%%%%%%%%%%%%%%%%%%%%%%%%%%%%%%%%%%%%%%
\begin{table*}
%\tiny
%\scriptsize 
\renewcommand{\tabcolsep}{0.1cm}
\begin{center}
\begin{tabular}{ccccccccccccccc} \hline
Obs. & Comp. & log U & log NH & $\rm{v_{out}}$ & $\Delta$C-stat & MC Sign. & $r_{launch}$ & $\dot{M}$ & $\dot{E}$ & $\dot{E}/L_{bol}$\\
& & & ($\rm{cm^{-2}}$) & (km/s) & & $\sigma$ & ($10^{16}$ cm) & ($M_{\odot}/yr$) & ($10^{42}$ erg/s) & \\  
\hline
XMM2000 & UFO\,1 & $\rm{-1.57\pm^{0.08}_{0.08}}$ & $\rm{21.08\pm^{0.08}_{0.10}}$ & 33,247$\pm$309 & 27 & 4.0 & $0.96^{+8.78}_{-0.86}$ & $0.018^{+0.203}_{-0.016}$ & $6.24
^{+72.0}_{-5.74}$ & ${8.42\times10^{-3}}^{+9.77\times10^{-2}}_{-7.74\times10^{-3}}$ \\
& UFO\,2  & $\rm{0.78\pm^{0.08}_{0.11}}$ & $\rm{21.63\pm^{0.14}_{0.15}}$ & 27,419$\pm$304 & 24 & 3.8 & $1.41^{+12.9}_{-1.27}$ & $0.077^{+1.01}_{-0.072}$ & $18.2
^{+246}_{-17.0}$ & ${2.46\times10^{-2}}^{+3.32\times10^{-1}}_{-2.29\times10^{-2}}$ \\
\hline
XMM2006 & UFO\,1 & $\rm{-0.17\pm^{0.17}_{0.11}}$ & $\rm{20.13\pm^{0.16}_{0.14}}$ & 20,787$\pm$299 & 32 & 4.8 & $2.45^{+22.7}_{-2.21}$ & $0.003^{+0.045}_{-0.003}$ & $0.437
^{+6.33}_{-0.41}$ & ${5.89\times10^{-4}}^{+8.53\times10^{-3}}_{-5.51\times10^{-4}}$ \\
& UFO\,2 & $\rm{-0.37\pm^{0.17}_{0.15}}$ & $\rm{19.81\pm^{0.15}_{0.18}}$ & 28,639$\pm$611 & 25 & 3.8 & $1.29^{+12.2}_{-1.17}$ & $0.001^{+0.016}_{-0.001}$ & $0.288
^{+4.25}_{-0.27}$ & ${3.89\times10^{-4}}^{+5.72\times10^{-3}}_{-3.66\times10^{-4}}$ \\
\hline
\end{tabular}
\end{center}
\caption{Best fit parameters of the ultra-fast outflow components detected in the RGS spectra of Mrk\,205 from the fit described in Section \ref{Results to the fit of the RGS spectra with PHASE}. Column 7 reports MonteCarlo significance (Section \ref{Monte Carlo Methods}). Properties in  columns 8-11 are derived in Section \ref{sec_disc_energectics}.}
\label{tab:UFOs parameters}
\end{table*}
%%%%%%%%%%%%%%%%%%%%%%%%%%%%%%%%%%%%%%%%%%%%%%%%%%%%%%%%%%%%%%%%%%%%%%%%%%%
%%%%%%%%%%%%%%%%%%%%%%%%%%%%%%%%%%%%%%%%%%%%%%%%%%%%%%%%%%%%%%%%%%%%%%%%%%%
%%%%%%%%%%%%%%%%%%%%%%%%%%%%%%%%%%%%%%%%%%%%%%%%%%%%%%%%%%%%%%%%%%%%%%%%%%%

\subsection{Physical model of the UFOs: PHASE} \label{Physical model for the UFOs}

To fit the RGS spectra of Mrk\,205 with a physical model, we used the photoionization code PHASE \citep{Krongold03}, which assumes absorbing gas in a plane-parallel geometry, and contemplates four parameters: the ionization parameter $U$, the equivalent hydrogen column density of the slab $N_H$, the outflow velocity of the wind $v_{out}$, and the microturbulence velocity of the medium $\sigma$. Note that, the ionization parameter defined in PHASE, is the  dimensionless ionization parameter $U=Q/4\pi R^{2}cn_{e}$, where $Q$ is the rate at which the source emits ionizing photons (photons/s), $c$ is the speed of light, $n_{e}$ is the electron density, and $R$ the is the distance of the gas from the X-ray source. 

To  calculate the ionization balance produced by the incoming ionizing radiation of the source, we constructed the spectral energy distribution (SED) of Mrk\,205 using data from the NASA/IPAC Extragalactic Database (NED) in the range of log($\nu$)$\sim$12-19 Htz and from the XMM2000 spectrum discussed here. 

\begin{figure*}
\begin{center}
\includegraphics[width=1.0\columnwidth]{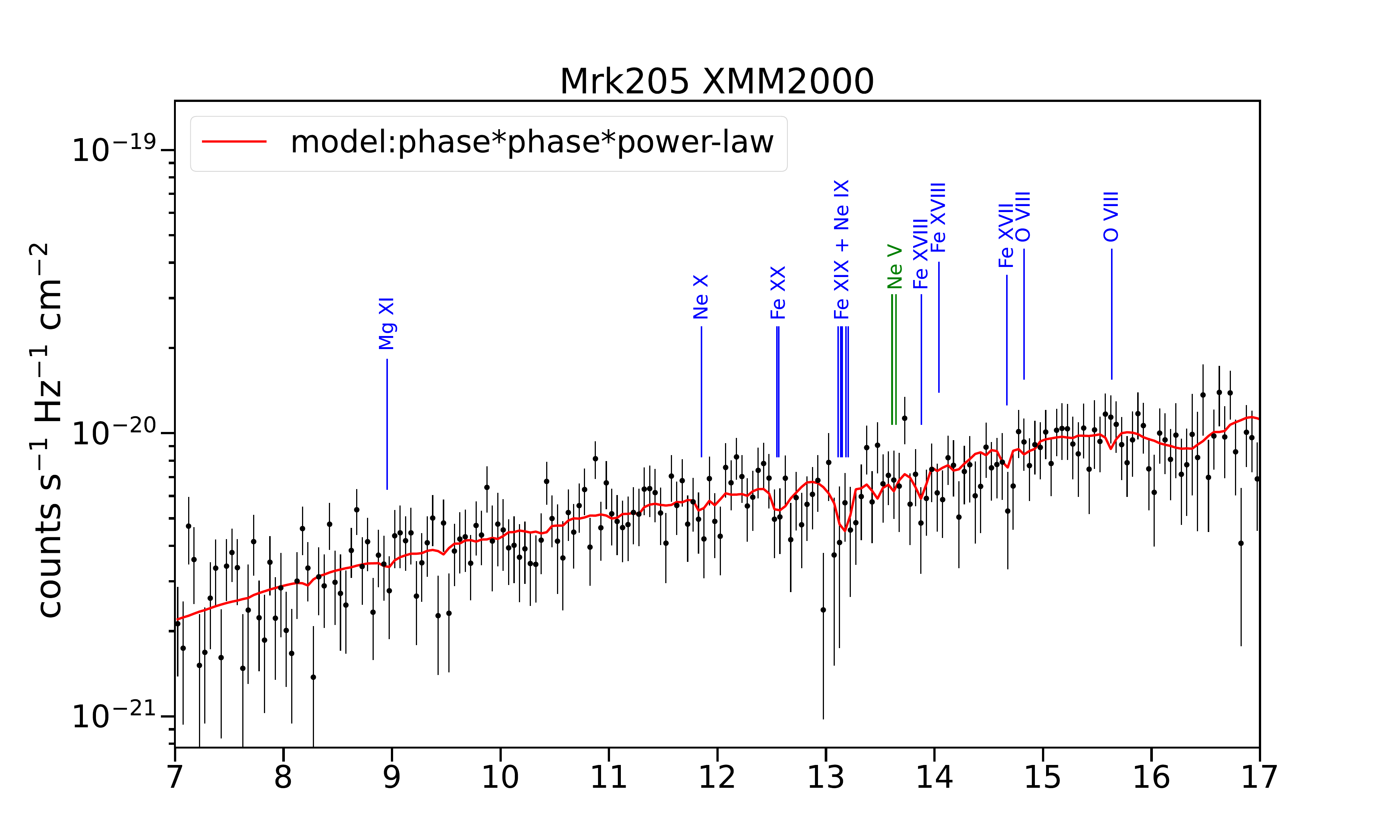}
\includegraphics[width=1.0\columnwidth]{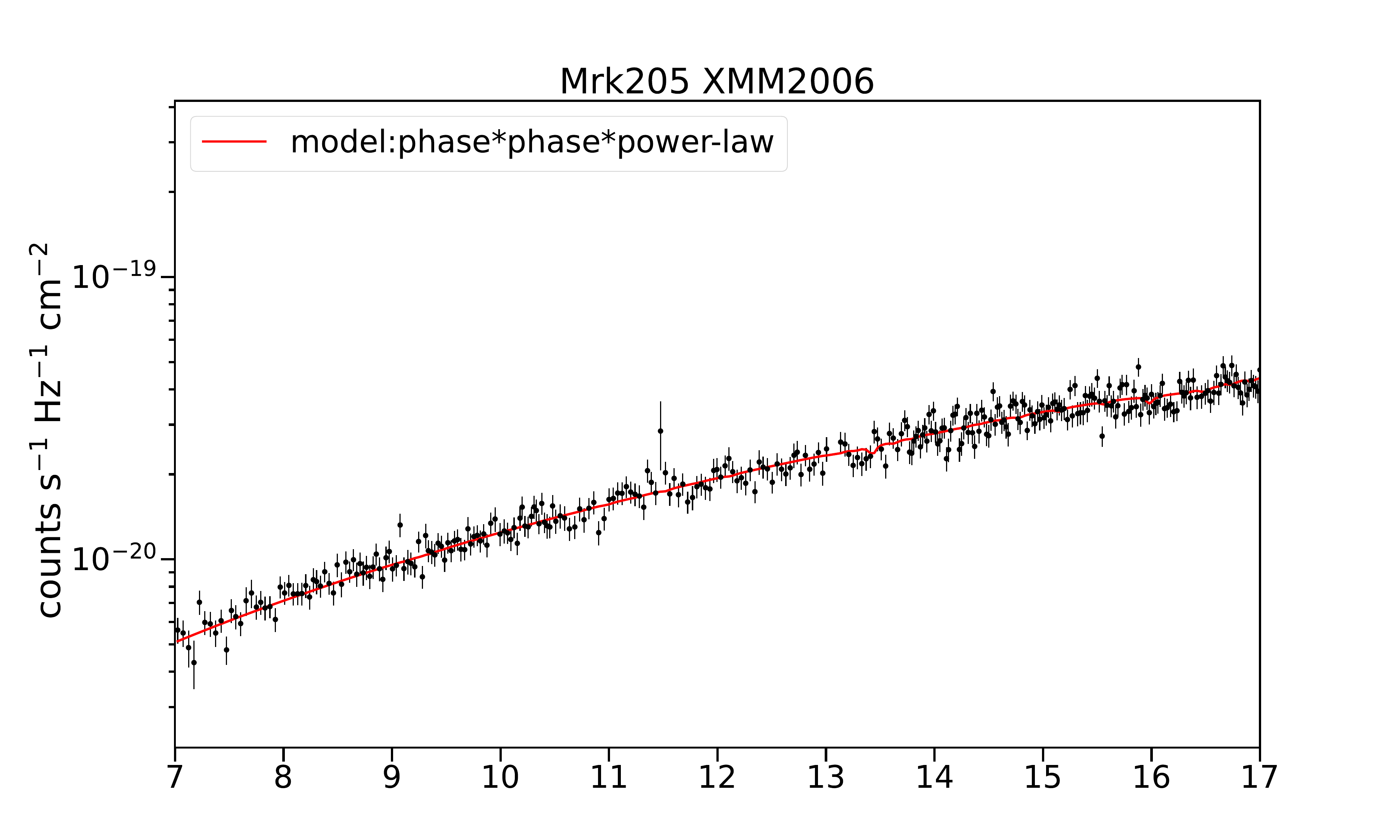}\\
\vspace{-0.3 cm}
\includegraphics[width=1.0\columnwidth]{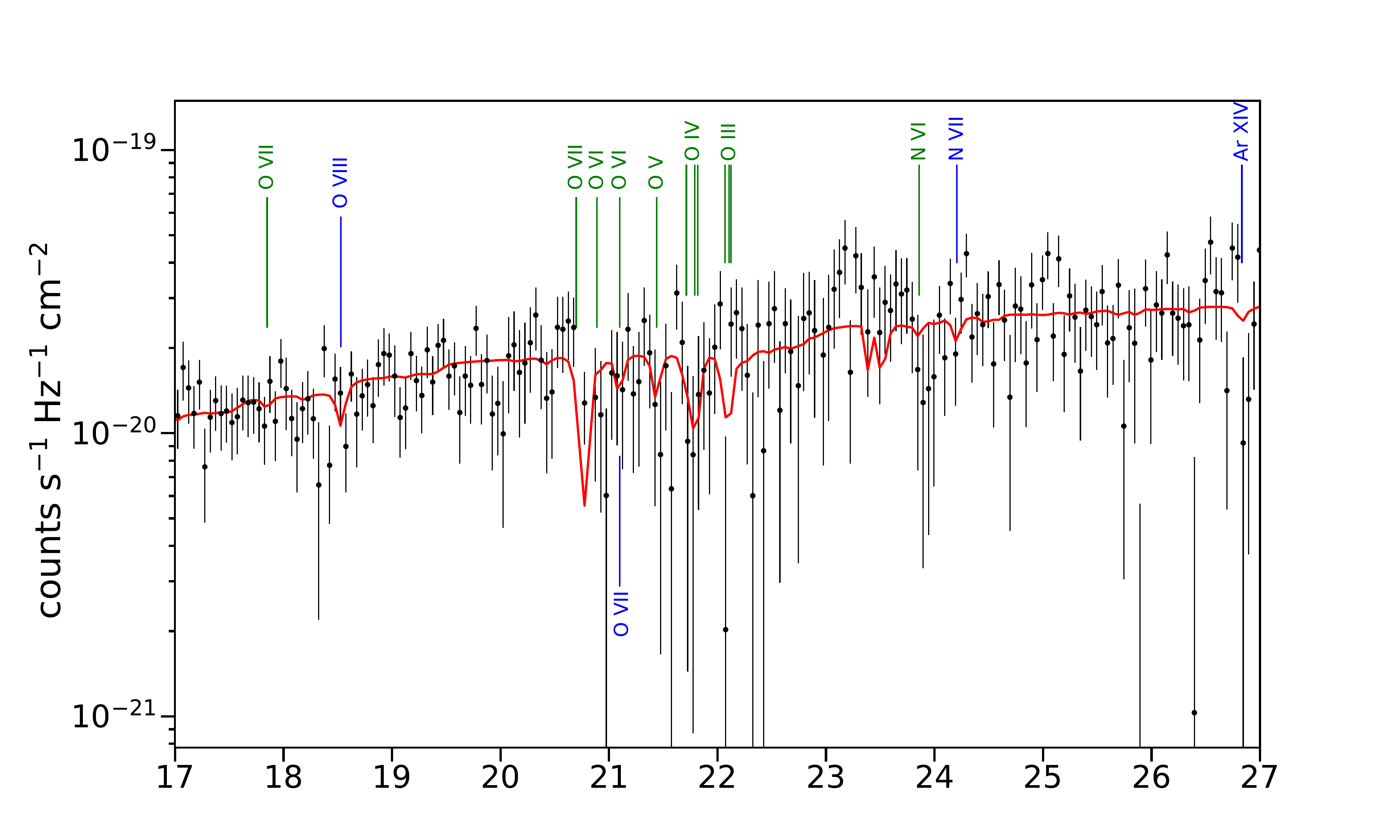}
\includegraphics[width=1.0\columnwidth]{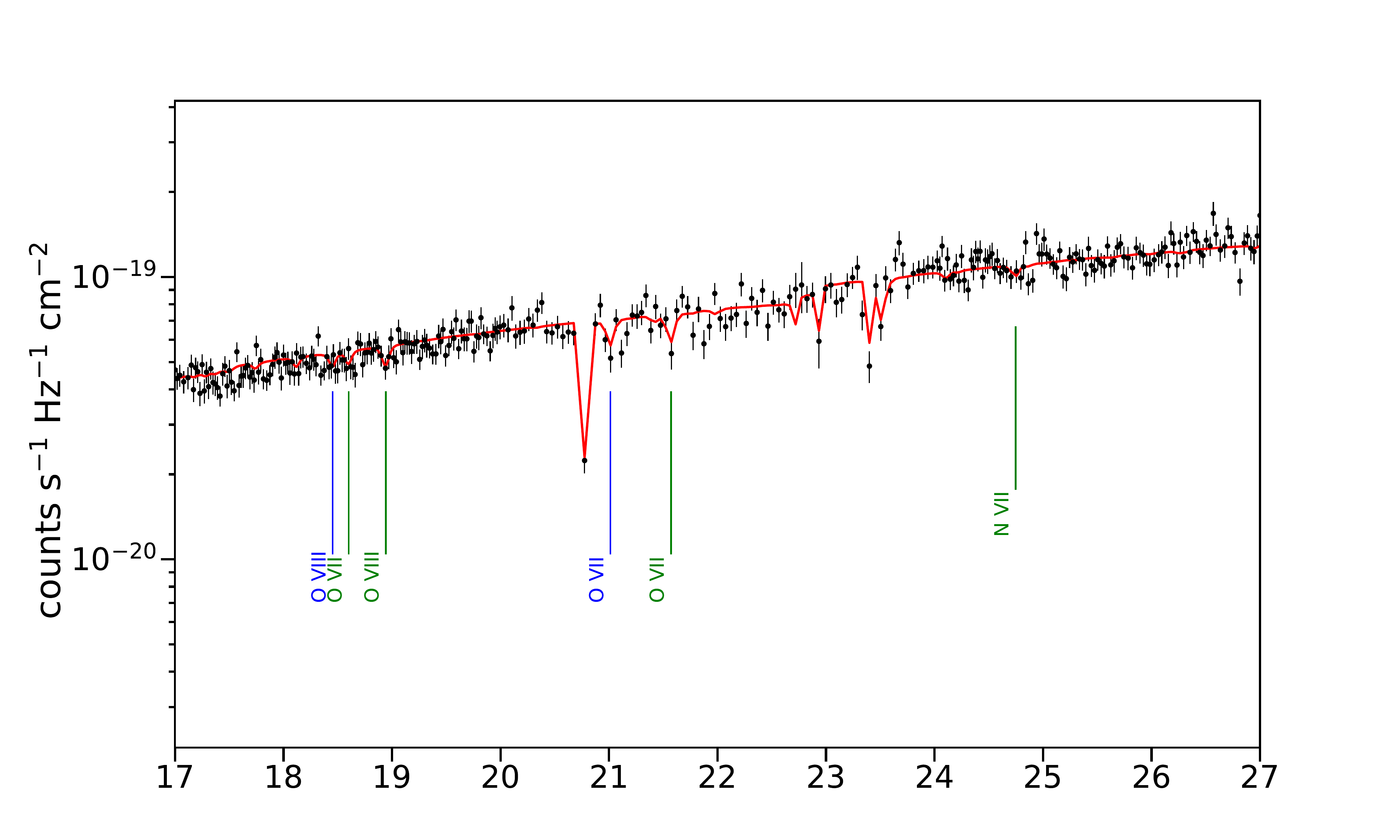}\\
\vspace{-0.3 cm}
\includegraphics[width=1.0\columnwidth]{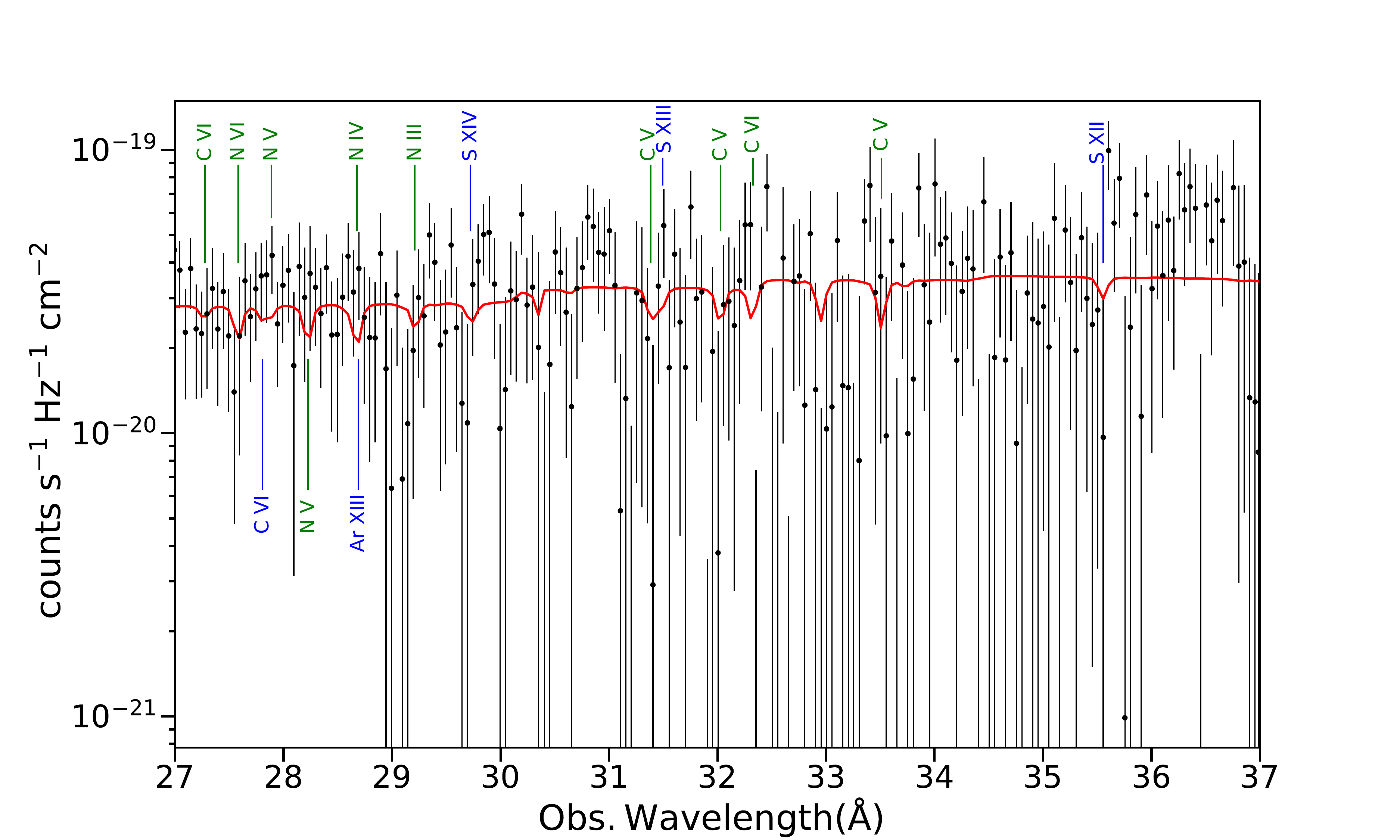}
\includegraphics[width=1.0\columnwidth]{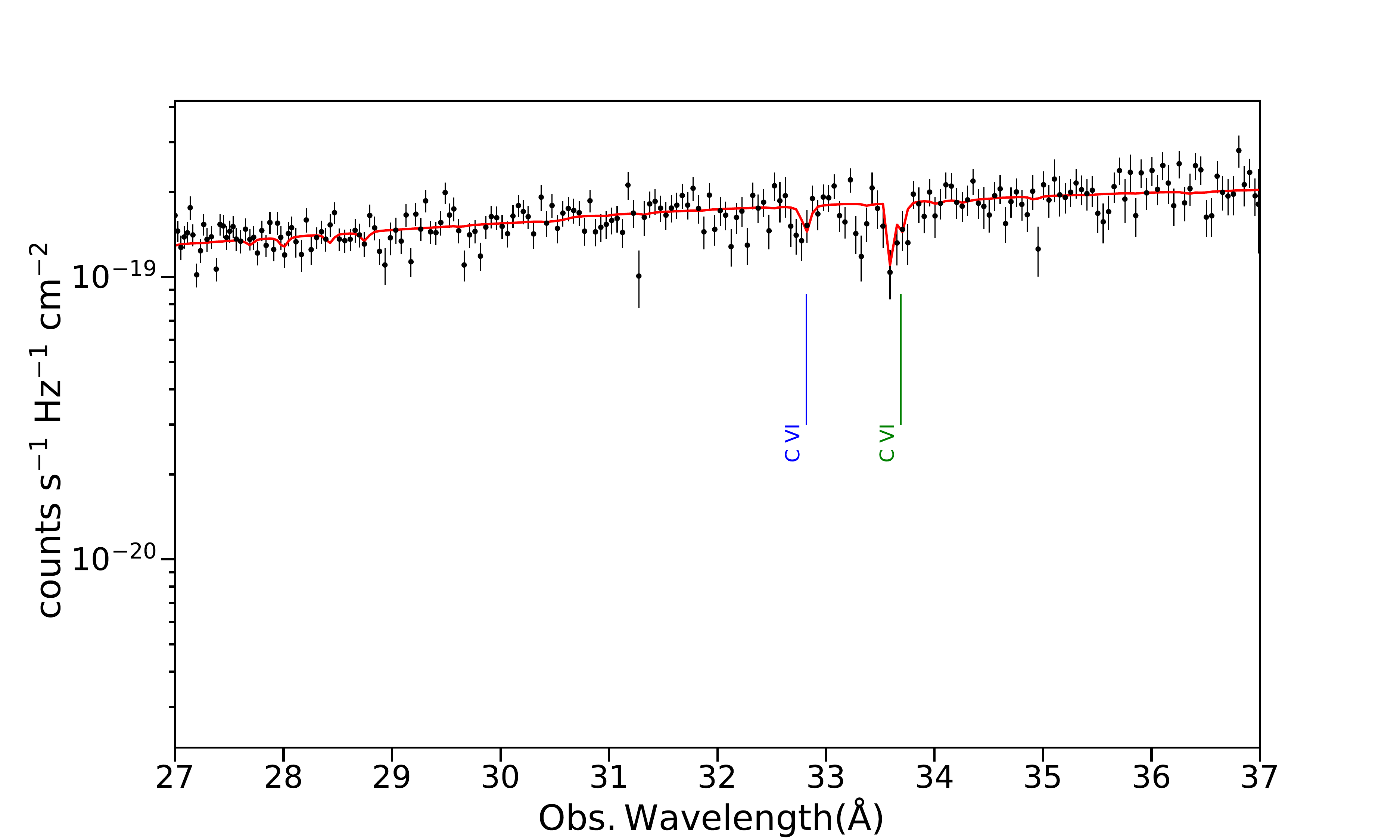}
\caption{RGS spectra of Mrk\,205 in the 7-37\,\r{A} range, fitted with a model including a power-law continuum and the two PHASE components accounting for the UFOs. The absorption features coincident with the UFO 1 are labeled in green, and those with the UFO 2 are marked in blue. The spectrum was binned for display purposes only. Left: XMM2000 observation. Right: XMM2006 observation.}
\label{fig:RGS PHASE}
\end{center}
\end{figure*}

To carry out the spectral fitting with PHASE, we fixed the turbulent velocity to 100\,km/s (because the lines are narrow and unresolved by RGS) and allowed the other three parameters of the model to vary, in addition to the power-law photon index. We fitted the spectra of the XMM2000 and XMM2006 observations of Mrk\,205 considering one, two, and three UFO systems, each modeled by a PHASE component. To determine if adding another UFO significantly improves the fit, we calculated the $\Delta$C-statistic corresponding to the improvement provided by the addition of each PHASE component. The final spectral fits are shown in Fig.\,\ref{fig:RGS PHASE}. We obtained the ionization parameter, the hydrogen column density, and the outflow velocity of the wind of each UFO component. The use of the self-consistent model allowed us to identify a large number of ionic transitions in the outflow, which are also reported in Fig.\,\ref{fig:RGS PHASE}. 

\subsubsection{Results of the RGS fit } \label{Results to the fit of the RGS spectra with PHASE}

We show in Tab.\,\ref{tab:UFOs parameters} (Cols. 3-5) the best-fit parameters of each UFO detected in the RGS spectra of Mrk\,205. We detected two UFO components in both epochs, XMM2000 and XMM2006. For the XMM2000 observation we found a $\Delta$C-stat of 27 and 24 (for three additional dof) when adding the UFO\,1 and UFO\,2 components, respectively. For the XMM2006 observation, the addition of UFO\,1 and UFO\,2 respectively returns $\Delta$C-stat of 32 and 25. See Col. 6 in Tab.\,\ref{tab:UFOs parameters}. We also marginally detected a third UFO component in both observations, with $\Delta$C-stat of 8 and 9 for the XMM2000 and XMM2006 observations, respectively. Note that since the  $\Delta$C-stat approximately follows a $\chi^2$ distribution for nested models and for the corresponding number of additional degrees of freedom \citep{Cash79, Kaastra17}, the two main UFO components are detected with >99\% confidence. For completeness, we also performed the spectral analysis on the three individual RGS 2006 observations, confirming the presence of the main UFO components in each dataset. The corresponding results are reported in Appendix\,\ref{appendix:XMM-Newton observations from 2006}.

According to our analysis, the UFO\,1 component of the XMM2000 observation reaches a velocity of $\sim$33,247\,km/s ($\sim$0.1\,c), while the UFO\,2 shows a lower velocity of $\sim$27,419\,km/s. The UFO\,1 shows a lower ionization and a lower column density compared to the UFO\,2. On the other hand, the UFO\,1 detected in the XMM2006 observation shows a lower velocity ($\sim$20,787\,km/s) compared to the UFO\,2 ($\sim$28,639\,km/s), but higher ionization and column density. We show in Fig.\,\ref{fig:RGS PHASE} the RGS spectra of Mrk\,205 fitted by the best-fit model. Note that the XMM2006 spectrum shows fewer absorption lines and less pronounced than the previous epoch.  On the other hand, the XMM2000 spectrum shows several absorption lines across the entire range, most of which correspond to UFO\,2 component, the one with the highest ionization.

\subsection{Monte Carlo Methods} \label{Monte Carlo Methods}

To provide an estimate of the statistical significance of the UFOs detected in XMM2000 and XMM2006 observations, we performed 5000 simulated spectra for each UFO with the \emph{fakeit} task in {\sc XSPEC} package. We restricted the simulations to the 7-37 \r{A} band, following the approach that was taken to model the UFOs in the RGS spectra of Mrk\,205. To test the significance of the UFO1 component we assumed for the synthetic data sets a baseline model composed by the continuum, modeled with an absorbed power-law. For the UFO2 component we included the continuum and the UFO1 component in the model as done in the spectral fitting of real data described in Section \ref{Results to the fit of the RGS spectra with PHASE}. For each simulated spectrum, all the free parameters of the additional PHASE component (ionization parameter, column density, and outflow velocity), together with the continuum parameters, were allowed to vary in order to obtain the best fit. Each simulated spectrum was folded through the same response matrix employed in the spectral fitting, with the same photon statistics as the real data set. In each simulated spectrum the improvement in C-statistics after adding a UFO component to the baseline model was measured and recorded, and we obtained the significance of the UFO components, reported in Tab.\,\ref{tab:UFOs parameters}.

%%%%%%%%%%%%%%%%%%%%%%%%%%%%%%%%%%%%%%%%%%%%%%%%%%%%%%%%%%%%%%%%%%%%%%%%%%%
%%%%%%%%%%%%%%%%%%%%%%%%%%%%%%%%%%%%%%%%%%%%%%%%%%%%%%%%%%%%%%%%%%%%%%%%%%%
%%%%%%%%%%%%%%%%%%%%%%%%%%%%%%%%%%%%%%%%%%%%%%%%%%%%%%%%%%%%%%%%%%%%%%%%%%%
%%%%%%%%%%%%%%%%%%%%%%%%%%%%%%%%%%%%%%%%%%%%%%%%%%%%%%%%%%%%%%%%%%%%%%%%%%%
%%%%%%%%%%%%%%%%%%%%%%%%%%%%%%%%%%%%%%%%%%%%%%%%%%%%%%%%%%%%%%%%%%%%%%%%%%%
%%%%%%%%%%%%%%%%%%%%%%%%%%%%%%%%%%%%%%%%%%%%%%%%%%%%%%%%%%%%%%%%%%%%%%%%%%%

%%%%%%%%%%%%%%%%%%%%%%%%%%%%%%%%%%%%%%%%%%%%%%%%%%%%%%%%%%%%%%%%%%%%%%%%%%%
%%%%%%%%%%%%%%%%%%%%%%%%%%%%%%%%%%%%%%%%%%%%%%%%%%%%%%%%%%%%%%%%%%%%%%%%%%%
%%%%%%%%%%%%%%%%%%%%%%%%%%%%%%%%%%%%%%%%%%%%%%%%%%%%%%%%%%%%%%%%%%%%%%%%%%%
\begin{table}
%\tiny
\scriptsize 
\begin{center}
\begin{tabular}{cccccccccccccccc} \hline
\multicolumn{3}{c}{UFO 1} & \multicolumn{3}{c}{UFO 2} \\
Ion & Obs. Wav & EW & Ion & Obs. Wav & EW \\
& (\r{A}) & (\r{A}) & & (\r{A}) & (\r{A}) \\
\hline
C\,V &  33.512  &   43.44 & C\,VI &  27.806 &    13.24  \\
C\,V &  32.029  &   33.09 & N\,VII &  24.206 &    22.18  \\
C\,V &  31.385  &   25.33 & O\,VII &  21.100 &    17.57  \\
C\,VI &  32.326  &   34.44 & O\,VIII &  18.529 &    35.09  \\
C\,VI &  27.277  &   11.34 & O\,VIII &  15.635 &    17.26  \\
N\,III &  29.227  &   11.72 & O\,VIII &  14.824 &    11.41 \\
N\,III &  29.211   &  12.59 & Ne\,IX &   13.135 &    13.59 \\
N\,IV &  28.678  &   36.52 & Ne\,X &   11.853 &    17.44 \\
N\,V &  28.227  &   33.47 & Mg\,XI &   8.956 &    10.04 \\
N\,V &  27.889  &   10.90 & S\,XII &  35.554 &    15.95 \\
N\,VI &  27.584  &   32.67 & S\,XIII &  31.494 &    14.88 \\
N\,VI &  23.858  &   12.82 & S\,XIV &  29.722 &    13.56 \\
O\,III &  22.125  &   27.81 & Ar\,XIII &   28.689 &    10.82 \\
O\,III &  22.106  &   24.34 & Ar\,XIV &   26.833 &    10.84 \\
O\,III &  22.068  &   21.22 & Fe\,XVII &   14.666 &    25.97 \\
O\,IV &  21.819  &   39.31 & Fe\,XVIII &   14.040 &    11.77 \\
O\,IV &  21.790  &   44.73 & Fe\,XVIII &   13.879 &    17.73 \\
O\,IV &  21.713  &   27.10 & Fe\,XVIII &   13.879 &    21.71 \\
O\,V &  21.439  &   44.55 & Fe\,XIX &   13.205 &    21.06 \\
O\,VI &  21.099  &   35.73 & Fe\,XIX &   13.184 &    14.65 \\
O\,VI &  20.889  &   16.46 & Fe\,XIX &   13.150 &    12.08 \\
O\,VII &  20.698  &   26.86 & Fe\,XIX &   13.112 &    10.12 \\
O\,VII &  17.849  &   10.78 & Fe\,XX &   12.566 &    10.46 \\
Ne\,V &  13.644  &   14.04 & Fe\,XX &   12.548 &    11.25 \\
Ne\,V &  13.609  &   10.78 \\
\end{tabular}
\end{center}
\caption{Atomic transitions in the winds detected in XMM2000 with an EW higher than 10 \r{A}, as described by the PHASE model.}
\label{tab:}
\end{table}
%%%%%%%%%%%%%%%%%%%%%%%%%%%%%%%%%%%%%%%%%%%%%%%%%%%%%%%%%%%%%%%%%%%%%%%%%%%
%%%%%%%%%%%%%%%%%%%%%%%%%%%%%%%%%%%%%%%%%%%%%%%%%%%%%%%%%%%%%%%%%%%%%%%%%%%
%%%%%%%%%%%%%%%%%%%%%%%%%%%%%%%%%%%%%%%%%%%%%%%%%%%%%%%%%%%%%%%%%%%%%%%%%%%

%%%%%%%%%%%%%%%%%%%%%%%%%%%%%%%%%%%%%%%%%%%%%%%%%%%%%%%%%%%%%%%%%%%%%%%%%%%
%%%%%%%%%%%%%%%%%%%%%%%%%%%%%%%%%%%%%%%%%%%%%%%%%%%%%%%%%%%%%%%%%%%%%%%%%%%
%%%%%%%%%%%%%%%%%%%%%%%%%%%%%%%%%%%%%%%%%%%%%%%%%%%%%%%%%%%%%%%%%%%%%%%%%%%
\begin{table}
%\tiny
\scriptsize 
\begin{center}
\begin{tabular}{cccccccccccccccc} \hline
\multicolumn{3}{c}{UFO 1} & \multicolumn{3}{c}{UFO 2} \\
Ion & Obs. Wav & EW & Ion & Obs. Wav & EW \\
& (\r{A}) & (\r{A}) & & (\r{A}) & (\r{A}) \\
\hline
C\,VI & 33.690 & 30.34 & C\,VI & 32.820 & 24.78 \\
N\,VII & 24.748 & 10.90 & O\,VII & 21.014 & 24.17 \\
O\,VII & 21.572 & 27.31 & O\,VIII & 18.454 & 12.29 \\
O\,VII & 18.602 & 11.14 & \\
O\,VIII & 18.943 & 19.92 & \\
\end{tabular}
\end{center}
\caption{Atomic transitions in the winds detected in XMM2006 with an EW higher than 10 \r{A}, as described by the PHASE model.}
\label{tab:}
\end{table}
%%%%%%%%%%%%%%%%%%%%%%%%%%%%%%%%%%%%%%%%%%%%%%%%%%%%%%%%%%%%%%%%%%%%%%%%%%%
%%%%%%%%%%%%%%%%%%%%%%%%%%%%%%%%%%%%%%%%%%%%%%%%%%%%%%%%%%%%%%%%%%%%%%%%%%%
%%%%%%%%%%%%%%%%%%%%%%%%%%%%%%%%%%%%%%%%%%%%%%%%%%%%%%%%%%%%%%%%%%%%%%%%%%%

\subsection{Spectral fitting of the CCD data} \label{Spectral fitting of the CCD data}

Motivated by the historical report of UFO in the Fe K band spectra of Mrk 205 by \cite{Tombesi10}, we now proceed with the investigation of the CCD spectra. We fitted the hard spectrum between 0.2-10 keV in the XMM-Newton observations. 
We designed a baseline model consisting of a power-law modified by the Galactic absorption to account for the X-ray continuum, the reflection model {\sc relxill} \citep{Dauser10, Garcia14} to account for the accretion disk reflection, and one emission line at 6.4\,keV to account for the narrow Fe\,K$\alpha$ emission line.
Then, we searched for absorption features associated with ultra-fast outflows. For this, we added a narrow Gaussian line ($\sigma$=0.01 keV) with negative intensity and free position. We computed the $\Delta\chi^2$ deviations from the best-fitting model and obtained the contour plots of the energy-intensity plane. This is the same procedure applied to the RGS spectra (see Sect.\,\ref{sec: Spectral fitting of the RGS data}).

\subsubsection{Results of the CCD spectral fit} \label{Results to the fit of the CCD spectra}

\begin{figure}
\begin{center}
\includegraphics[width=1.0\columnwidth]{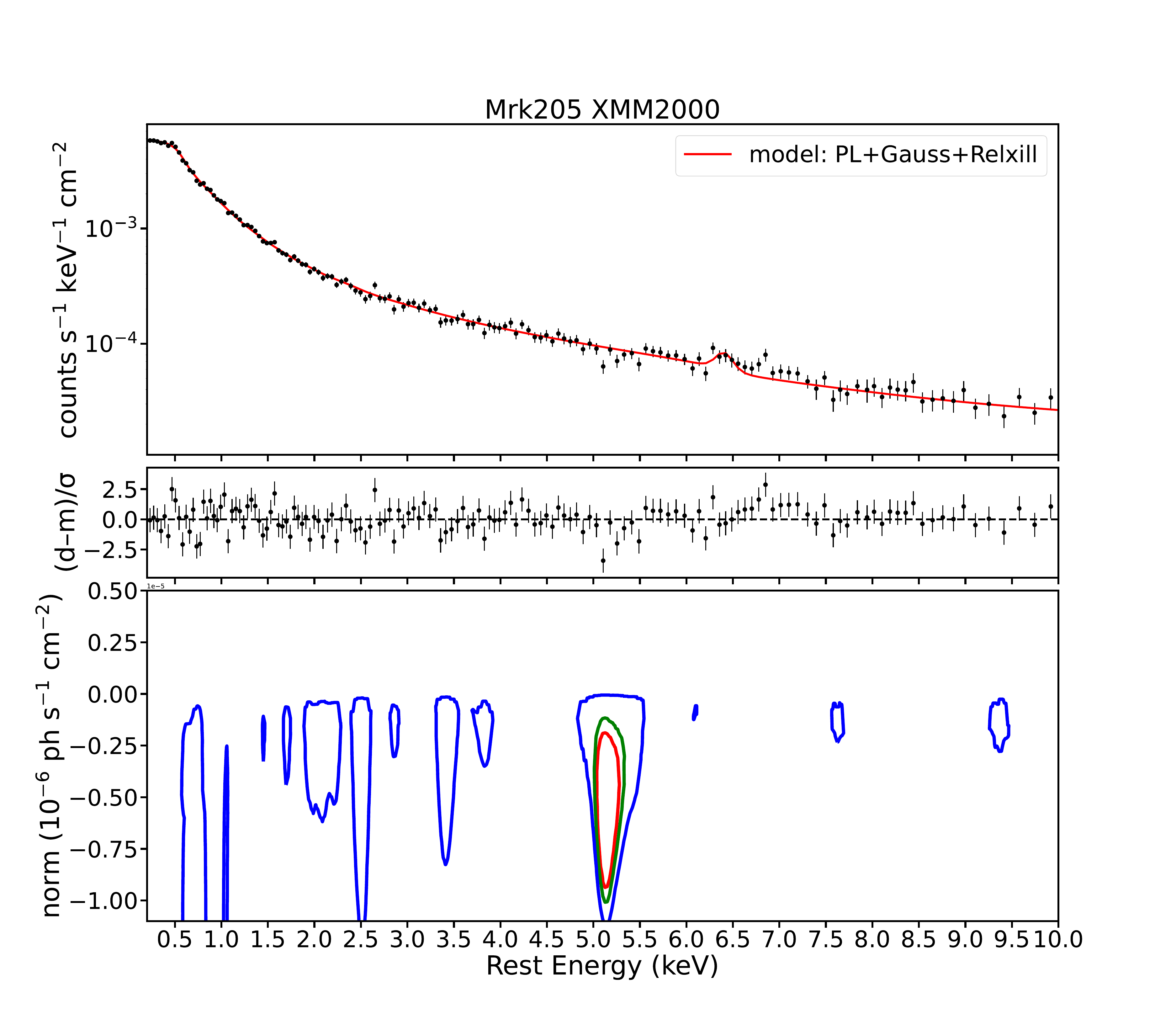 }
\caption{Top: Best fitting-model to the XMM2000 spectrum of Mrk\,205 in the 0.2-10\,keV range. Middle: Residuals of the best fitting-model. Bottom: Two-dimensional $\Delta\chi^2$ contours for the resulting line search. Red, green, and blue contours correspond to confidence levels of 68$\%$, 90$\%$, and 99$\%$, respectively.}
\label{fig:CCD spectra and contours}
\end{center}
\end{figure}

Table.\,\ref{tab:line search CCD} reports the parameters of the absorption features detected in the CCD spectrum of the XMM2000 observation of Mrk\,205. We identified a tentative absorption feature at rest-frame energy of $\sim$ 7.63 keV (see left panel in Fig.\,\ref{fig:CCD spectra and contours}). Given its low statitical significance ($\sim$1.9$\sigma$) and the precense of other spectral residuals, we consider this detection only tentative. Therefore, we do not use this feature as evidence of an UFO, but only as a consitency check whit the feature previosuly detected by \cite{Tombesi10}. In any case, if the $\sim$ 7.63 keV feature is associated with Fe\,XXVI Ly$\alpha$, this would correspond to an outflow velocity of $\sim$0.095\,c.

As to XMM2006, since the individual and combined pn spectra are fully consistent with no absorption, we omitted the spectral parameters of the XMM2006 in Table\,\ref{tab:line search CCD}. However, for completeness, we report results of the spectral analysis on the three individual XMM pn 2006 observations  in Appendix\,\ref{appendix:XMM-Newton observations from 2006}, which are consistent with the null detection of \cite{Tombesi10}.
 
%%%%%%%%%%%%%%%%%%%%%%%%%%%%%%%%%%%%%%%%%%%%%%%%%%%%%%%%%%%%%%%%%%%%%%%%%%%
%%%%%%%%%%%%%%%%%%%%%%%%%%%%%%%%%%%%%%%%%%%%%%%%%%%%%%%%%%%%%%%%%%%%%%%%%%%
%%%%%%%%%%%%%%%%%%%%%%%%%%%%%%%%%%%%%%%%%%%%%%%%%%%%%%%%%%%%%%%%%%%%%%%%%%%
\begin{table*}
%\tiny
%\scriptsize 
\renewcommand{\tabcolsep}{0.1cm}
\begin{center}
\begin{tabular}{c|c|c|ccccc|ccccc} \hline
& PL & FeK$\alpha$ & \multicolumn{5}{c}{Reflection ({\sc relxill})} & \multicolumn{5}{c}{Absorption features}  \\
\hline
Obs. & $\Gamma$ & Rest. Ene & Index & a & Incl. & $\rm{log(\xi)}$ & $\rm{Z_{Fe}}$ & Rest Ene. & EW & Sign. & $\Delta\chi^2$ & $v_{out}^{FeXXVI}$ \\
& & (keV) & & & (deg) & & & (keV) & (eV) & ($\sigma$) & & (c) \\  \hline
XMM2000 & $\rm{1.79\pm^{0.02}_{0.02}}$ & $\rm{6.39\pm^{0.04}_{0.05}}$ & <4.81 & $\rm{0.86\pm^{0.09}_{0.15}}$ & $\rm{34\pm^{2}_{11}}$ & $\rm{2.34\pm^{0.06}_{0.06}}$ & $\rm{4.45\pm^{0.30}_{0.44}}$ & $\rm{7.63\pm^{0.07}_{0.07}}$ & $\rm{-49\pm^{26}_{26}}$ & 1.9 & 3 & $\rm{0.0908\pm0.0091}$ \\ 
\hline
\end{tabular}
\end{center}
\caption{Best-fit parameters of the baseline model, including the detected absorption features, used to fit the CCD spectra of Mrk 205. Baseline model includes the power-law continuum, the FeK$\alpha$ emission line, and the accretion disk reflection. The {\sc relxill} parameters are Index: index emissivity of the disk; a: spin of the black hole; Incl: inclination angle; $\rm{\xi}$: ionization parameter (in erg cm $\rm s^{-1}$); $\rm{Z_{Fe}}$: iron abundance (relative to solar).}
\label{tab:line search CCD}
\end{table*}
%%%%%%%%%%%%%%%%%%%%%%%%%%%%%%%%%%%%%%%%%%%%%%%%%%%%%%%%%%%%%%%%%%%%%%%%%%%
%%%%%%%%%%%%%%%%%%%%%%%%%%%%%%%%%%%%%%%%%%%%%%%%%%%%%%%%%%%%%%%%%%%%%%%%%%%
%%%%%%%%%%%%%%%%%%%%%%%%%%%%%%%%%%%%%%%%%%%%%%%%%%%%%%%%%%%%%%%%%%%%%%%%%%%

%%%%%%%%%%%%%%%%%%%%%%%%%%%%%%%%%%%%%%%%%%%%%%%%%%%%%%%%%%%%%%%%%%%%%%%%%%%
%%%%%%%%%%%%%%%%%%%%%%%%%%%%%%%%%%%%%%%%%%%%%%%%%%%%%%%%%%%%%%%%%%%%%%%%%%%
%%%%%%%%%%%%%%%%%%%%%%%%%%%%%%%%%%%%%%%%%%%%%%%%%%%%%%%%%%%%%%%%%%%%%%%%%%%
%%%%%%%%%%%%%%%%%%%%%%%%%%%%%%%%%%%%%%%%%%%%%%%%%%%%%%%%%%%%%%%%%%%%%%%%%%%
%%%%%%%%%%%%%%%%%%%%%%%%%%%%%%%%%%%%%%%%%%%%%%%%%%%%%%%%%%%%%%%%%%%%%%%%%%%
%%%%%%%%%%%%%%%%%%%%%%%%%%%%%%%%%%%%%%%%%%%%%%%%%%%%%%%%%%%%%%%%%%%%%%%%%%%

\section{Discussion}\label{sec:Discussion}

In this work, we studied the multi-epoch X-ray spectra of the low-luminosity Quasar Mrk\,205, observed by  \emph{XMM}-Newton in 2000 and 2006. For the 2006 epoch,  we combined the three available observations to maximize the signal-to-noise ratio. The \emph{XMM}-Newton data were analysed using the high-resolution RGS apectra. We also found marginal evidence for one UFO component in the EPIC-pn spectrum of the 2000 observation. According to our analysis of the RGS spectra,  Mrk\,205 exhibits multiphase ultra-fast outflows at both epochs, clearly showing two distinct components in the high-resolution data.

\subsection{Comparison with the previously detected UFO} \label{sec: Discussion CCD spectra}

Mrk\,205 has been previously studied by \cite{Tombesi10} in the context of ultra-fast outflows as part of a sample of objects. These authors analyzed the \emph{XMM}-Newton EPIC-pn observations of the same two epochs studied in this work. They reported  no features associated with ultra-fast outflows in the 2006 observations (XMM2006-A and XMM2006-C), however they detected a narrow absorption feature in the spectrum of 2000. According to their analysis, the absorption line at 7.7$\pm$0.03 keV, if identified with Fe\,XXVI Ly$\alpha$ resonant absorption, implies an outflow velocity of 0.1$\pm$0.004c.
Our results of the CCD data analysis from the same observation are broadly in agreement with the results obtained by \cite{Tombesi10}. Indeed, our reported  detection of absorption at 7.63$\pm$0.07 keV suggests an outflow velocity of 0.095$\pm$0.c. 

We do not attempt to calculate the energetics of the ultra-fast outflow feature detected in the Fe\,K band due to this absorption feature seen in the CCD spectra of \emph{XMM}-Newton have low statistical significance. Therefore, we consider the Fe K band detection as tentative, although it certainly provide a useful consistency check with the findings in the seminal paper by \cite{Tombesi10}, and use it only as complementary evidence to the higher-confidence results obtained from the RGS analysis.
Deeper exposures of Mrk 205 with substantially improved signal-to-noise ratios are essential to confirm the tentative feature reported here and to better constrain their physical parameters.

\subsection{Multi-component episodic UFOs in Mrk\,205}

According to our results about the physical parameters of the UFOs detected in the RGS spectra, the UFO\,1 component detected in the two epochs shows clear differences in ionization parameter, column density, and outflow velocity (see Tab.\,\ref{tab:UFOs parameters}), suggesting that the components observed in 2000 and 2006 trace in fact, different UFOs.

The UFO\,2 component present in XMM2000 and XMMM2006 shows different ionization parameter and column density, but fairly similar outflow velocities, which could suggest an origin in a persistent wind. However,  the ionization state and the column density of UFO2 decrease from 2000 to 2006 (see Tab.\ref{tab:UFOs parameters}), while the flux increases (see Tab.\ref{tab:Observational_parameters}), suggesting that they are different UFOs, since higher ionization of the gas would be expected as a response  to the higher luminosity of the source. To investigate whether UFO2 has expanded in the time elapsed between the \emph{XMM}-Newton observations of 2000 and 2006, we refer to the definitions of $U=Q/4\pi R^{2}cn_{e}$. In this definition $Q$ is the rate at which the source emits ionizing photons (photons/s), $c$ is the speed of light, $n_{e}$ is the gas electron density, and $R$  is the distance of the gas from the X-ray source; In this framework, $N_H=n_e\Delta r$, where $N_H$ is the gas column density, and $\Delta r$ is the thickness of the absorbing layer. We calculated the change in $n_{e}r^{2}$ between the \emph{XMM}-Newton epochs, finding that it must have increased $\sim$36 times from 2000 to 2006. In addition, we find that $N_H$ decreases $\sim$ 66 times between the first and second observations. According to this analysis, if  UFO\,2 component is the same wind observed at different epochs, the thickness of the expanding layer must be extremely thin, which suggests that they are two distinct components. Furthermore, assuming the observed outflow velocity, the absorber would travel approximately $5.8\times10^{17}$ cm during the six years elapsed between the XMM-Newton observations, more than one order of magnitude larger than its estimated launching radius. Therefore, it is unlikely that the absorber detected in 2006 corresponds to the same physical structure observed in 2000. These considerations indicate that the UFO components detected in the RGS spectra of Mrk\,205 are likely  different components, suggesting outflow episodes rather than persistent winds in Mrk\,205.

On the other hand, the detection of the two UFO components in both epochs showing different parameters of ionization, column density, and velocity suggests the presence of multiphase outflows in Mrk\,205. Similar results have been found for other objects. For instance, \cite{Longinotti15} reported the first multi-component ultra-fast outflow detected in X-ray high-resolution spectroscopic data in the NLSy1 galaxy IRAS\,17020+4544. \cite{Parker20} find a scenario consistent with an outflowing wind radially stratified, observed in the X-ray data of the quasar IRAS\,13349+2438 with \emph{XMM}-Newton and \emph{NuSTAR}. \cite{Krongold21} analyzed the \emph{XMM}-Newton spectra of Mrk\,1044, finding evidence for a multiphase ultra-fast outflow in the RGS spectrum, consisting of four different absorbing components.

Recent results obtained with \emph{XRISM} have confirmed this complex structure of the ultra-fast outflows. \cite{Xiang25} find that in the Fe\,K band, X-ray spectra of NGC\,4151 require as many as six wind absorption components, indicating a stratified, multiphase wind. \cite{Xu25} and \cite{Mizumoto26} studied observations of XRISM, \emph{XMM}-Newton and \emph{NuSTAR} of PDS\,456 and PG\,1211+143, respectively, resolving multiple clumpy wind components in the Fe\,K band in both sources. 

In order to investigate if the UFO components detected in XMM2000 and XMM2006 observations are in pressure equilibrium we calculated the relative pressure of the absorbers assuming photoionization equilibrium, using the relation $P \propto T/U$. Temperatures were derived from a grid of photoionization models computed with CLOUDY. For XMM2000 we obtained $P \sim {8.51\times10^{5}}$ and $P \sim {6.93\times10^{4}}$ for UFO\,1 and UFO\,2 components, respectively. On the other hand, we obtained $P \sim {1.21\times10^{5}}$ and $P \sim {1.33\times10^{5}}$ for UFO\,1 and UFO\,2 components, respectively from XMM2006. The similar values of P of UFO\,1 and UFO\,2 components of XMM2006 indicate that they are consistent with pressure equilibrium, suggesting that the absorbers may belong to a multi-phase medium. In contrast, for XMM2000, the two UFO components show a discrepancy in P of more than an order of magnitude, indicating that they are not in pressure equilibrium. This suggests that the absorbers likely arise from physically distinct regions or represent transient structures.

\begin{figure}
\begin{center}
\includegraphics[width=0.9\columnwidth]{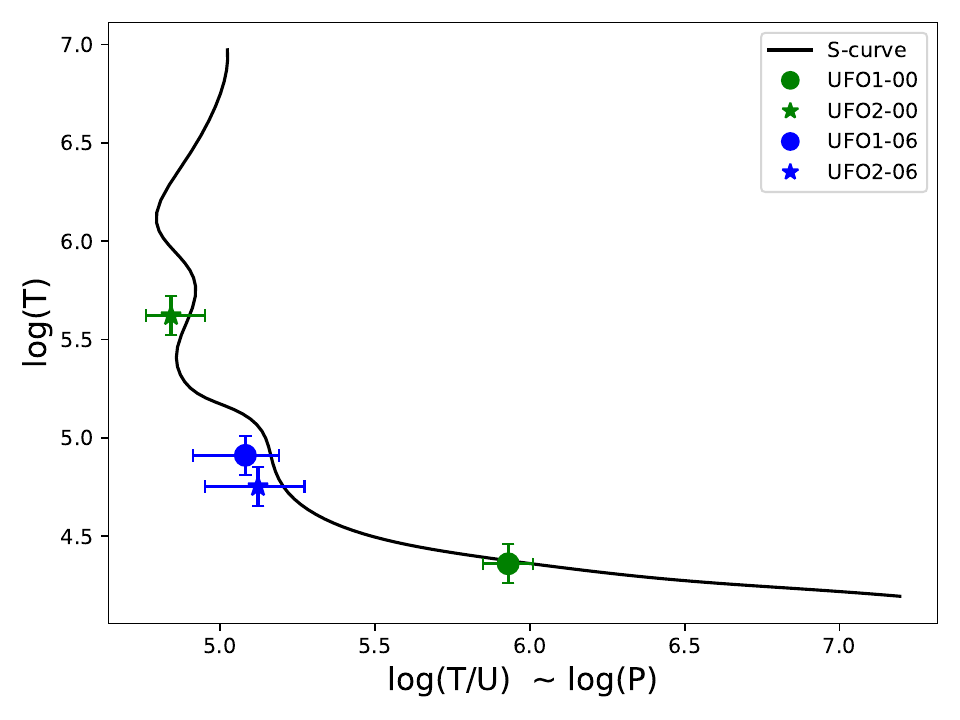}
\caption{Plot of the Thermal stability curve (S-curve) of Mrk 205 (black line). UFO components detected in the RGS spectra of XMM-Newton are overplotted.}
\label{fig:Scurve}
\end{center}
\end{figure}

We also explored whether the absorbers could be explained within a Radiation Pressure Confinement (RPC) scenario \citep[eg.][]{Stern14} , in which the gas is expected to form a continuous distribution of ionization states under approximate pressure balance. According to our results about equilibrium pressure, the UFO components detected in XMM2006 are compatible with the framework of RPC, while the UFO components in XMM2000 show that the RPC scenario is not sufficient to explain the behavior of these absorbers. In addition, we constructed the thermal stability curve (S-curve) of Mrk\,205 adopting the source spectral energy distribution, which shows the equilibrium temperature as a function of the pressure proxy T/U. Fig.\,\ref{fig:Scurve} shows that the UFO components in XMM2006 are located at similar values of T/U, indicating comparable gas pressures, and lie on thermally stable branches of the curve. This supports the presence of a multi-phase medium in pressure equilibrium, consistent with expectations from RPC. On the other hand, UFO components detected in XMM2000 show different values of T/U, implying different gas pressures. Moreover, the UFO\,2 component detected in XMM2000 is located near a thermally unstable region of the curve. These results indicate that the absorbers are unlikely to form a stable multi-phase structure.

%%%%%%%%%%%%%%%%%%%%%%%%%%%%%%%%%%%%%%%%%%%%%%%%%%%%%%%%%%%%%%%%%%%%%%%%%%%
%%%%%%%%%%%%%%%%%%%%%%%%%%%%%%%%%%%%%%%%%%%%%%%%%%%%%%%%%%%%%%%%%%%%%%%%%%%
%%%%%%%%%%%%%%%%%%%%%%%%%%%%%%%%%%%%%%%%%%%%%%%%%%%%%%%%%%%%%%%%%%%%%%%%%%%
%%%%%%%%%%%%%%%%%%%%%%%%%%%%%%%%%%%%%%%%%%%%%%%%%%%%%%%%%%%%%%%%%%%%%%%%%%%
%%%%%%%%%%%%%%%%%%%%%%%%%%%%%%%%%%%%%%%%%%%%%%%%%%%%%%%%%%%%%%%%%%%%%%%%%%%
%%%%%%%%%%%%%%%%%%%%%%%%%%%%%%%%%%%%%%%%%%%%%%%%%%%%%%%%%%%%%%%%%%%%%%%%%%%

 \subsection{Driving mechanism}

We also investigated the possible driving mechanism of the UFOs detected in the RGS spectra of Mrk\,205. According to \cite{King15}, in the radiatively driven scenario, energy-conserving outflows require a conserved kinetic power \.E$_{k}\propto v_{out}^{3}/ \xi$, therefore,  $v_{out}\propto \xi^{1/3}$. On the other hand, in the MHD driven scenario, the relation is between $v_{out}\propto \xi^{1/2}$ and $v_{out}\propto \xi^{0.7}$ \citep[see][and references therein]{Xu25}. To test these relations, we converted the dimensionless ionization parameter $U$ given by PHASE to the ionization parameter $\xi$. Note that the ionization parameters $U$ and $\xi$ are related through $\xi = kU$, where $k \sim 20-40$ for typical AGN \citep{Krolik81, Netzer96, Blustin05, Mehdipour16}. We used a value for $k$ of 30. 

\begin{figure}
\begin{center}
\includegraphics[width=0.9\columnwidth]{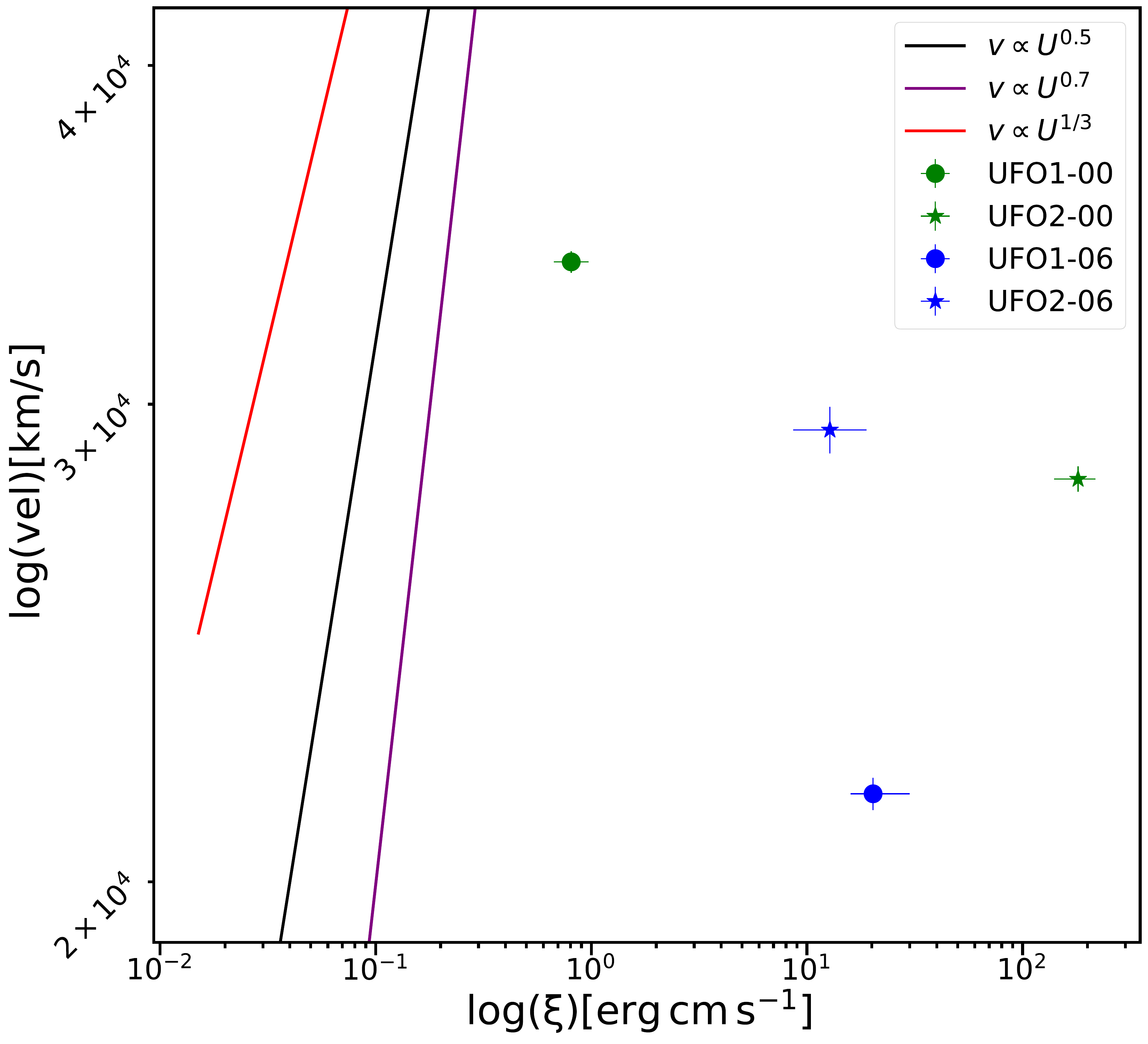}
\caption{Plot of outflow velocity versus ionization parameter of the four UFOs detected in the RGS spectra of Mrk\,205. Red line represents the relationship predicted by the radiatively driven scenario, black and purple lines represent the relationships predicted by the MHD driven.}
\label{fig:Driving mechanism}
\end{center}
\end{figure}

Radiatively driven UFOs are associated with highly accreting AGN, while less highly accreting AGN are expected to be dominated by the MHD driven scenario \citep{Fukumura17,Xu25}. According to the inefficient accretion observed in Mrk\,205, MHD is expected to be the driving mechanism of the UFOs present in this source. We show in Fig.\,\ref{fig:Driving mechanism} the expected relations from both driving mechanism scenarios and the UFOs detected in Mrk\,205.  According to the plot, no UFO component is consistent with the radiatively driven scenario or MHD predictions, suggesting a possible complex driving mechanism, probably with the contribution of both launching scenarios.  

%%%%%%%%%%%%%%%%%%%%%%%%%%%%%%%%%%%%%%%%%%%%%%%%%%%%%%%%%%%%%%%%%%%%%%%%%%%
%%%%%%%%%%%%%%%%%%%%%%%%%%%%%%%%%%%%%%%%%%%%%%%%%%%%%%%%%%%%%%%%%%%%%%%%%%%
%%%%%%%%%%%%%%%%%%%%%%%%%%%%%%%%%%%%%%%%%%%%%%%%%%%%%%%%%%%%%%%%%%%%%%%%%%%
%%%%%%%%%%%%%%%%%%%%%%%%%%%%%%%%%%%%%%%%%%%%%%%%%%%%%%%%%%%%%%%%%%%%%%%%%%%
%%%%%%%%%%%%%%%%%%%%%%%%%%%%%%%%%%%%%%%%%%%%%%%%%%%%%%%%%%%%%%%%%%%%%%%%%%%
%%%%%%%%%%%%%%%%%%%%%%%%%%%%%%%%%%%%%%%%%%%%%%%%%%%%%%%%%%%%%%%%%%%%%%%%%%%
 
\subsection{Energetics of the Ultra-fast outflows and implications for feedback}
\label{sec_disc_energectics}

In order to investigate the contribution to the host galaxy feedback of the UFOs detected in the RGS spectra of Mrk\,205, we estimated the mass and energy carried by each component. To do this, we started by calculating the minimum launching radius of each UFO component. For this, we assume that the outflow velocity is higher than or equal to the escape velocity at the launch radius $r_{launch}=2GM/v_{out}^2$. To calculate this radius, we used the SMBH mass reported in the literature, $log(M_{BH}/M_{\odot})=8.6\pm1.0$ \citep{Wandel86}. Then, we calculate the mass outflow rate, $\dot{M}_{out}=4\pi\mu m_{p}r N_{H} v_{out} C_f$, where $m_p$ and $\mu$ correspond to the mass of the proton and to the mean atomic mass per particle ($\mu$=1.4), $C_f$ to the covering fraction of the outflow, and $N_H$, $v_{out}$ and $r$ to the column density, the observed velocity, and the launching radius of the outflow, respectively. From the mass outflow rate we calculate the energy outflow rate, or kinetic luminosity through the expression $\dot{E}=\frac{1}{2}\dot M_{out}v_{out}^2C_f$. Finally, we calculated the ratio of the kinetic energy to the bolometric luminosity, $\dot{E}/L_{bol}$.  

 Tab.\,\ref{tab:UFOs parameters} (Cols. 8-11) reports the launching radius, the mass outflow rate, the energy outflow rate, and the ratio of the kinetic energy to the bolometric luminosity obtained for each UFO component detected in the RGS spectra of Mrk\,205. The estimated minimum launching radius  is consistent with an  accretion disk origin. The mass outflow rate remains low for all RGS UFOs (<0.1 $M_{\odot}/yr$). We found that the UFOs detected in the XMM2006 spectrum exhibit lower mass outflow rate and energy outflow rate compared to the UFOs detected in the XMM2000 spectrum. Finally, the ratio $\dot{E}/L_{bol}$ is also lower for the XMM2006 UFO components.  

According to \cite{Hopkins10}, the minimum power required for a wind that transfers energy outward to produce feedback is $\geq$0.5$\%$ of the bolometric luminosity. 
For the XMM2000 observation, we obtain $\dot{E}/L_{bol}$ $\approx$ 0.8$\%$ $C_{f}$ and $\dot{E}/L_{bol}$ $\approx$ 2.5$\%$ $C_{f}$ for  UFO\,1 and UFO\,2, respectively. On the other hand, UFO\,1 and UFO\,2 in XMM2006 yield $\dot{E}/L_{bol}$ $\approx$ 0.06$\%$ $C_{f}$ and $\dot{E}/L_{bol}$ $\approx$ 0.04$\%$ $C_{f}$, respectively. The minimum power required is satisfied for $C_{f} \geq 0.63$ and $C_{f} \geq 0.2$ for UFO\,1 and UFO\,2 in XMM2000, respectively. The derived $\dot{E}/L_{bol}$ during the XMM2006 epoch remains below the minimum feedback requirement even for $C_{f}=1$. Note that the values obtained of $\dot{E}/L_{bol}$ bear caveats associated with the SMBH mass that goes into the estimation of the launch radius of the wind. Due to the large uncertainty in the SMBH mass estimate \citep{Tombesi10}, the ratio of $\dot{E}/L_{bol}$ should be taken with care. On the other hand, we note that for outflows with velocities of $\gtrsim$0.05c, relativistic effects can introduce corrections to the observed column density \citep{Luminari20, Luminari24}, potentially increasing the inferred kinetic energy for the fastest components.  Although a detailed relativistic treatment is beyond the scope of this work.

The marked difference of $\dot{E}/L_{bol}$ between the two epochs indicates that the energy output of the wind may vary significantly on timescales of years. This could be explained since Mrk\,205 exhibits inefficient accretion ($\lambda_{Edd}$ $\sim$ 0.03), possibly due to a truncated disc as previously suggested by \cite{Laha19} to account for the absence of a broad FeK emission line. In this scenario, the differences derived in the ratio $\dot{E}/L_{bol}$ could be related to changes in the geometry of the inner accretion disc, where changes in the truncation radius could affect the wind launching conditions, producing less efficient outflow phases.

\section*{Acknowledgements}

This work was supported by UNAM Posdoctoral Program (POSDOC). ALL and CV-C acknowledge support from DGAPA-PAPIIT IA103625. Lara-DI acknowledges support from Secretaría de Ciencia, Humanidades, Tecnología e Innovación (SECIHTI) through the Estancias Posdoctorales por México Convocatoria 2025 (Estancia Posdoctoral Iniciales) grant program.

%%%%%%%%%%%%%%%%%%%%%%%%%%%%%%%%%%%%%%%%%%%%%%%%%%
\section*{Data Availability}

The data underlying this article are publicly available from the \emph{XMM}-Newton Science Archive.

%%%%%%%%%%%%%%%%%%%%%%%%%%%%%%%%%%%%%%%%%%%%%%%%%%

%%%%%%%%%%%%%%%%% APPENDICES %%%%%%%%%%%%%%%%%%%%%

\appendix

%%%%%%%%%%%%%%%%%%%%%%%%%%%%%%%%%%%%%%%%%%%%%%%%%%
%\section{Appendix}

\section{\emph{XMM}-Newton observations from 2006} \label{appendix:XMM-Newton observations from 2006}

In order to investigate the presence of ultra-fast outflows in the individual observations from 2006, we analyzed the three \emph{XMM}-Newton RGS and CCD spectra of Mrk\,205. First, we used the PHASE code to fit the RGS spectra in the range of 7-37 \r{A}. We show in Tab.\,\ref{tab:UFOS_RGS} the parameters obtained for each UFO component detected in the individual observations. We confirmed the presence of the two UFO components in all three individual observations. The best-fit ionization parameters and column densities are consistent within their uncertainties, while modest velocity differences are observed among the individual observations, which may be associated with a clumpy or inhomogeneous wind structure. Due to limitation of the current S/N of individual RGS spectra, these possibilities would be better tested with future deeper observations.

We then repeat the procedure for the CCD data: we fitted the pn spectra of Mrk\,205 in the 0.2-10\,keV range to test for the presence of the UFO features described in Section \ref{Results to the fit of the CCD spectra}. We show in Tab.\,\ref{tab:line_search_CCD} the parameters obtained for the UFO features in each pn spectrum of  the XMM2006 data. As mentioned several times throughout the paper, the CCD tests are included to provide a useful comparison with previous results.

%%%%%%%%%%%%%%%%%%%%%%%%%%%%%%%%%%%%%%%%%%%%%%%%%%%%%%%%%%%%%%%%%%%%%%%%%%%
%%%%%%%%%%%%%%%%%%%%%%%%%%%%%%%%%%%%%%%%%%%%%%%%%%%%%%%%%%%%%%%%%%%%%%%%%%%
%%%%%%%%%%%%%%%%%%%%%%%%%%%%%%%%%%%%%%%%%%%%%%%%%%%%%%%%%%%%%%%%%%%%%%%%%%%
\begin{table*}
%\tiny
%\scriptsize 
%\renewcommand{\tabcolsep}{0.1cm}
\begin{center}
\begin{tabular}{ccccccccccccc} \hline
Obs. & Component & log U & log NH & $\rm{v_{out}}$ & $\Delta$C-stat \\ 
& & & ($\rm{cm^{-2}}$) & (km/s) & \\
\hline
XMM2006-A & Comp 1 & $\rm{-0.20\pm^{0.45}_{0.64}}$ & $\rm{20.17\pm^{0.41}_{0.62}}$ & 23,053$\pm$297 & 10 \\ 
 & Comp 2 & $\rm{-0.47\pm^{0.30}_{0.47}}$ & $\rm{19.90\pm^{0.22}_{0.26}}$ & 34,453$\pm$287 & 8 \\
\hline
XMM2006-B & Comp 2 & $\rm{-0.44\pm^{0.40}_{0.26}}$ & $\rm{20.03\pm^{0.20}_{0.19}}$ & 22,453$\pm$297 & 12 \\ 
 & Comp 1 & $\rm{-1.26\pm^{0.66}_{0.17}}$ & $\rm{20.07\pm^{0.15}_{0.19}}$ & 29,653$\pm$291 & 19 \\
\hline
XMM2006-C & Comp 2 & $\rm{-0.5\pm^{0.25}_{0.22}}$ & $\rm{20.08\pm^{0.17}_{0.23}}$ & 26,053$\pm$294 & 11 \\ 
 & Comp 1 & $\rm{-2.0\pm^{0.31}_{0.14}}$ & $\rm{19.99\pm^{0.32}_{0.19}}$ & 31,453$\pm$290 & 18 \\ 
\hline
\end{tabular}
\end{center}
\caption{Best fit parameters of the ultra-fast outflow components detected in the RGS spectra of the individual observations of Mrk\,205 from 2006.}
\label{tab:UFOS_RGS}
\end{table*}
%%%%%%%%%%%%%%%%%%%%%%%%%%%%%%%%%%%%%%%%%%%%%%%%%%%%%%%%%%%%%%%%%%%%%%%%%%%
%%%%%%%%%%%%%%%%%%%%%%%%%%%%%%%%%%%%%%%%%%%%%%%%%%%%%%%%%%%%%%%%%%%%%%%%%%%
%%%%%%%%%%%%%%%%%%%%%%%%%%%%%%%%%%%%%%%%%%%%%%%%%%%%%%%%%%%%%%%%%%%%%%%%%%%

%%%%%%%%%%%%%%%%%%%%%%%%%%%%%%%%%%%%%%%%%%%%%%%%%%%%%%%%%%%%%%%%%%%%%%%%%%%
%%%%%%%%%%%%%%%%%%%%%%%%%%%%%%%%%%%%%%%%%%%%%%%%%%%%%%%%%%%%%%%%%%%%%%%%%%%
%%%%%%%%%%%%%%%%%%%%%%%%%%%%%%%%%%%%%%%%%%%%%%%%%%%%%%%%%%%%%%%%%%%%%%%%%%%
\begin{table*}
%\tiny
%\scriptsize 
%\renewcommand{\tabcolsep}{0.1cm}
\begin{center}
\begin{tabular}{cc|ccc|cccccccc} \hline
Obs. & Rest Ene. & EW & Sign. & $\chi^2$/d.o.f. & $\Delta$ $\chi^2$ \\
& (keV) & (eV) & ($\sigma$) & &  \\  \hline
XMM2006-A & $\rm{8.63\pm^{0.05}_{0.05}}$ & $\rm{-42\pm^{21}_{22}}$ & 2 & 153.22/116 & 4 \\ \hline
XMM2006-B & $\rm{9.87\pm^{0.08}_{0.08}}$ & $\rm{-62\pm^{34}_{34}}$ & 1.8 & 143.90/110 & 3  \\
 & $\rm{7.57\pm^{0.05}_{0.06}}$ & $\rm{-28\pm^{18}_{18}}$ & 1.5 & 144.94/110 & 2 \\ \hline
XMM2006-C & $\rm{9.17\pm^{0.05}_{0.05}}$ & $\rm{-36\pm^{19}_{19}}$ & 1.9 & 148.55/119 & 3 \\
\hline
\end{tabular}
\end{center}
\caption{Parameters of the absorption features detected in the CCD spectra of the individual observations of Mrk\,205 from 2006.}
\label{tab:line_search_CCD}
\end{table*}
%%%%%%%%%%%%%%%%%%%%%%%%%%%%%%%%%%%%%%%%%%%%%%%%%%%%%%%%%%%%%%%%%%%%%%%%%%%
%%%%%%%%%%%%%%%%%%%%%%%%%%%%%%%%%%%%%%%%%%%%%%%%%%%%%%%%%%%%%%%%%%%%%%%%%%%
%%%%%%%%%%%%%%%%%%%%%%%%%%%%%%%%%%%%%%%%%%%%%%%%%%%%%%%%%%%%%%%%%%%%%%%%%%%

\section{Additional test on XMM2000 RGS spectrum} \label{Additional test on XMM2000 RGS spectrum}

In order to investigate whether the best-fit model derived from the XMM2006 spectrum can adequately reproduce the absorption features observed in XMM2000 observation, we performed additional tests by applying the XMM2006 best-fit model to the XMM2000 spectrum while allowing different subsets of parameters to vary. Specifically, we explored cases in which only the continuum photon index was re-fitted, as well as cases in which the photon index and the ionization parameter or the photon index and the column density were allowed to vary. The resulting fit statistics are reported in Tab.\,\ref{tab:RGS tests}. In all cases, the fit remained statistically worse than the best-fit solution obtained directly for the XMM2000 spectrum. This result supports the conclusion that the absorber properties differ significantly between the two epochs.

%%%%%%%%%%%%%%%%%%%%%%%%%%%%%%%%%%%%%%%%%%%%%%%%%%%%%%%%%%%%%%%%%%%%%%%%%%%
%%%%%%%%%%%%%%%%%%%%%%%%%%%%%%%%%%%%%%%%%%%%%%%%%%%%%%%%%%%%%%%%%%%%%%%%%%%
%%%%%%%%%%%%%%%%%%%%%%%%%%%%%%%%%%%%%%%%%%%%%%%%%%%%%%%%%%%%%%%%%%%%%%%%%%%
\begin{table*}
%\tiny
%\scriptsize 
%\renewcommand{\tabcolsep}{0.1cm}
\begin{center}
\begin{tabular}{cccccccccccc} \hline
Test & C-stat/d.o.f. & $\Delta$C \\ \hline
XMM2000 model & 3128.72/2973 & 0 \\
XMM2006 model & 3374.62/2980 & +246 \\
$\Gamma$ & 3173.00/2979 & +44 \\
$\Gamma$ + U & 3169.48/2977 & +41 \\
$\Gamma$ + nH & 3162.43/2977 & +34 \\
\hline
\end{tabular}
\end{center}
\caption{C-stat/d.o.f. and $\Delta$C values resulting when applying the XMM2006 best-fit model to the XMM2000 spectrum while allowing different subsets of parameters to vary.}
\label{tab:RGS tests}
\end{table*}
%%%%%%%%%%%%%%%%%%%%%%%%%%%%%%%%%%%%%%%%%%%%%%%%%%%%%%%%%%%%%%%%%%%%%%%%%%%
%%%%%%%%%%%%%%%%%%%%%%%%%%%%%%%%%%%%%%%%%%%%%%%%%%%%%%%%%%%%%%%%%%%%%%%%%%%
%%%%%%%%%%%%%%%%%%%%%%%%%%%%%%%%%%%%%%%%%%%%%%%%%%%%%%%%%%%%%%%%%%%%%%%%%%%

\section{CGM of NGC\,4319}

\begin{figure}
\begin{center}
\includegraphics[width=0.9\columnwidth]{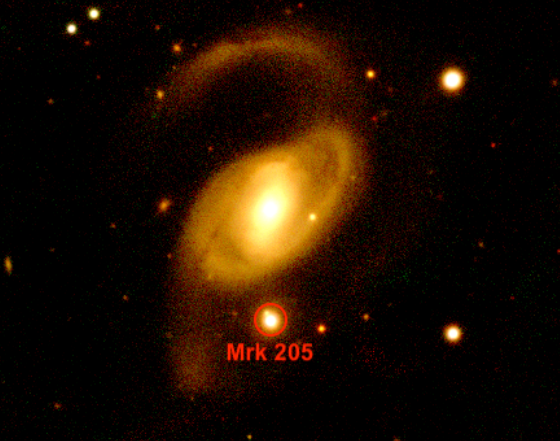}
\caption{Optical image of Mrk\,205 (marked with the red circle), with the foreground spiral galaxy NGC\,4319}
\label{fig:Mrk205_NGC4319}
\end{center}
\end{figure}

Because Mrk\,205 lies along a line of sight where the galaxy NGC\,4319 is located in the foreground (see Fig.\,\ref{fig:Mrk205_NGC4319}), we investigated whether any absorption line produced by the circumgalactic medium (CGM) of NGC\,4319 appears in the analyzed spectra of Mrk\,205 and whether it might contribute to the detected UFOs. To do this, we first considered typical CGM absorption lines (N\,VI, N\,VII, O\,VI, O\,VII, O\,VIII, Ne\,IX, and Ne\,X), and redshifted them to the systemic velocity of NGC\,4319 to find their offset. To find their expected position, we used the equation $\lambda_{exp}=\lambda_{rest}(1+z_{NGC4319})$, where $\lambda_{exp}$ is the expected wavelength, $\lambda_{rest}$ is the rest-frame wavelength, and $z_{NGC4319}$ is the redshift of NGC\,4319. We then compared the predicted positions with the list of lines detected in our blind line search. We found that only the N\,VI line that could correspond to the CGM of NGC\,4319 was detected in our line search, at 29.928 \r{A}. However, we only detected it in the XMM2006 observation, while no CGM line matched any detection in the XMM2000 observation.

To test for a possible CGM component from NGC\,4319 in the Mrk\,205 spectra, we added a PHASE component to our best-fit model, which includes the two UFO components. This component was fixed at the redshift of NGC\,4319, with its velocity set to 0 km/s (since CGM gas is not expected to have high velocities), and we allowed to vary the ionization parameter and column density. We found that adding a CGM component did not improve the fit, resulting in a $\Delta$C-stat of 1. Among the transitions found due to this component, we again found the N\,VI transition, with an equivalent width of 1.73 \r{A}.

Finally, typical CGM absorption features are extremely weak, with typical equivalent widths of m\r{A}, and are usually detected only in staked spectra from several observations \citep[see eg.][]{Lara24}. Note that the absorption lines corresponding to the UFOs we have detected have equivalent widths >10\r{A}. Therefore, we do not expect any CGM line from NGC\,4319 to be misidentified as a UFO in the spectra of Mrk\,205.

%%%%%%%%%%%%%%%%%%%%%%%%%%%%%%%%%%%%%%%%%%%%%%%%%%%%%%%%%%%%%%%%%%%%%%%%%%%
%%%%%%%%%%%%%%%%%%%%%%%%%%%%%%%%%%%%%%%%%%%%%%%%%%%%%%%%%%%%%%%%%%%%%%%%%%%
%%%%%%%%%%%%%%%%%%%%%%%%%%%%%%%%%%%%%%%%%%%%%%%%%%%%%%%%%%%%%%%%%%%%%%%%%%%
%%%%%%%%%%%%%%%%%%%%%%%%%%%%%%%%%%%%%%%%%%%%%%%%%%%%%%%%%%%%%%%%%%%%%%%%%%%
%%%%%%%%%%%%%%%%%%%%%%%%%%%%%%%%%%%%%%%%%%%%%%%%%%%%%%%%%%%%%%%%%%%%%%%%%%%
%%%%%%%%%%%%%%%%%%%%%%%%%%%%%%%%%%%%%%%%%%%%%%%%%%%%%%%%%%%%%%%%%%%%%%%%%%%

%%%%%%%%%%%%%%%%%%%%%%%%%%%%%%%%%%%%%%%%%%%%%%%%%%

% Don't change these lines
\bsp	% typesetting comment
\label{lastpage}
\end{document}